\documentclass[preprint,12pt]{elsarticle}

\usepackage{hyperref}
\usepackage{amssymb}
\usepackage{amsmath}
\usepackage{geometry}
\usepackage{graphicx}
\usepackage{cleveref}
\usepackage{subfigure}
\usepackage{bm}
\usepackage{longtable}
\usepackage{multirow}
\usepackage{float}
\usepackage{fullpage}
\usepackage{booktabs}
\usepackage[numbers]{natbib}
\usepackage{color}
\usepackage{xcolor}
\usepackage{diagbox}
\usepackage{longtable}
\usepackage[figuresright]{rotating}

\usepackage{lineno}

\makeatletter
\def\ps@pprintTitle{%
	\let\@oddhead\@empty
	\let\@evenhead\@empty
	\let\@oddfoot\@empty
	\let\@evenfoot\@oddfoot
}
\makeatother

\begin{document}

\begin{frontmatter}



  \title{Optimal energy harvesting efficiency from vortex-induced
  	vibration of a circular cylinder under flow}


\author[labela]{Peng Han\corref{cor}}
\cortext[cor]{Corresponding author.}
\ead{hanpeng@mail.tsinghua.edu.cn}

\author[labelb]{Qiaogao Huang}
\author[labelb]{Guang Pan}

\author[labelb]{Denghui Qin}


\author[labelc]{Wei Wang}

\author[labeld]{Rodolfo T. Gonçalves}

\author[labele,labelf]{Jisheng Zhao\corref{cor2}}
\cortext[cor2]{Corresponding author.}
\ead{Jisheng.Zhao@monash.edu}



\address[labela]{AML, Department of Engineering Mechanics, Tsinghua University, Beijing, 100084, P.R.China}

\address[labelb]{School of Marine Science and Technology, Northwestern Polytechnical University, Xi'an 710072, P.R.China}


\address[labelc]{Department of Mechanical Engineering, The Hong Kong Polytechnic University, Hong Kong, China}

\address[labeld]{Ocean Space Planning Laboratory (OSPL), Department of Systems Innovation, The University of Tokyo, Tokyo 106-0032, Japan}

\address[labele]{Fluids Laboratory for Aeronautical and Industrial Research (FLAIR), Department of Mechanical and Aerospace Engineering, Monash University, Clayton 3800, Australia}

\address[labelf]{School of Engineering and Information Technology, University of New South Wales, Canberra, ACT
10 2600, Australia}

\begin{abstract}
This work applies a combined approach a reduced-order model (ROM)
together with experiments and direct numerical simulations to
investigate the optimal efficiency of fluid-flow energy harvesting
from transverse vortex-induced vibration (VIV) of a circular
cylinder. High-resolution efficiency maps were predicted over wide
ranges of flow reduced velocities and structural damping ratios, and
the maximum efficiency and optimal settings of damping ratio and
reduced velocity were then examined for different mass ratios and
Reynolds numbers. Efficiencies predicted by the ROM were also
validated against either experiments or direct simulations. The present work
indicates that: (i) the maximum efficiency is controlled by both the incoming
reduced velocity and the product of mass ratio and structural
damping ratio, which is similar to the maximum amplitude of VIV;
(ii) the maximum efficiency at a relatively high Reynolds number
($Re \approx 6 \times 10^3$) in subcritical regime is higher than
that of a low Reynolds number ($Re = 150$) in laminar regime; (iii)
the energy harvesting efficiency from VIV of a circular cylinder
with a low mass ratio is more robust than that with a high mass
ratio. This finding suggests that the VIV harvester performs better
in water than in air.

\end{abstract}


\end{frontmatter}

\section{Introduction}
\label{Sec1}
When an elastic or elastically mounted body is submitted to a transverse flow
(for instance, wind or ocean currents), it may vibrate under the
fluctuating fluid forces exerted by vortex
shedding, 
and then the body vibration in turn affects the surrounding flow field
\cite{Paidoussis2010}. This typical fluid-structure interaction (FSI)
phenomenon is called vortex-induced vibration (VIV). VIV can be widely
found in engineering applications and natural lives, such as bridges
and buildings (civil engineering), aircraft wings (aerospace
engineering), offshore oil risers and mooring lines (ocean
engineering), and vibrating leaves and plants in nature
\cite{Blevins2001}. The circular cylinder has been employed as the
standard geometry for investigating VIV, as the circular shape is
always symmetrical to the incoming free-stream flow, and the
fluid-structure system can therefore be immune to the influence of
other instabilities, such as galloping
\citep[see][]{Barrero-Gil2009a,zhao2014fsi,
	zhao2018Dsection}. The fundamental research interest
and practical importance of flow-induced vibration has motivated a
large body of work on modelling, characterising, and predicting the
vibration response and vortex shedding modes, well documented in the
reviews of Refs \citep{Williamson2004,Sarpkaya2004} and the books of Refs \cite{Paidoussis2010,Blevins2001}.

One of the most important features of VIV is {\em synchronisation}
(also know as {\em ``lock-in''}), where both the vortex shedding
frequency and the body vibration frequency are close/equal to the
natural frequency, leading to large-amplitude oscillations
\cite{Williamson2004}. Usually, VIV, particularly in lock-in, is
treated as a kind of destructive, high-energy, and harmful phenomenon
to the structural safety and thus the primary limiting factor in
structural designs, because the structural vibration can potentially
lead to structural fatigue and failures \cite{Kandasamy2016,Wang2020}.
However, in the past decade, researchers have found that the
structural vibration can absorb considerable fluid kinetic energy, and
thus it is regarded as a promising potential for energy harvesting
\cite{Bernitsas2008,RostamiRSER2017,Wang2020AE,Wang2021APL}. It
has been experimentally and numerically shown that fluid-flow energy
harvesting from VIV has many advantages when compared with other
conventional techniques such as rotary turbines; for instance, apart
from their lower cost to operate and maintain, VIV energy converters
can work over a broad range of incoming flow velocities and even at
very-low-speed currents, such as rivers and shallow waters
\cite{Bernitsas2008,Ma2016,Ding2015}.

To date, extensive computational fluid dynamics (CFD) and experimental
studies have been conducted on enhancing the efficiency of VIV energy
harvesters under different structural properties and flow conditions,
including the structural damping ratio, surface roughness, and
incoming flow velocity
\citep[e.g.][]{Ma2016,Ding2015,WangECM2019,SotiJFS2018,HanOE,zhao2022prf,LeeOE}.
It has been indicated that the maximum efficiency point always locates
within the ``lock-in'' regime, and it is sensitive to the structural
damping, incoming flow velocity, and Reynolds numbers. However,
because of the limitations of experimental setup and the high costs of
CFD tools, particularly Direct Numerical Simulation (DNS), it is
almost impossible to conduct a detailed optimisation study with a
comprehensive parametric space covering wide ranges of the structural
properties and fluid conditions. Thus, reduced-order models, with
extremely low costs, are effective alternatives to compute and
optimise energy conversion efficiency from VIV, despite its less
accuracy \citep[see][]{Barrero-Gil2012, Abdelkefi2012,Lai2021MSSP}.

In this paper, taking the advantages of different research methods for
VIV, we present a comprehensive study, by combining a reduced-order model, a quasi-DNS-based FSI solver,
and experiments, to better understand and optimise the VIV energy conversion efficiency. We will
show that the reduced-order modelling is of cost-efficiency to
effectively estimate the peak amplitude response and energy harvesting
efficiency of VIV of a circular cylinder, as compared with our direct numerical and experimental results. Of interest, our reduced-order
modelling results will show that the global maximum energy harvesting
efficiency is barely affected by mass ratio (the ratio of the
oscillating mass to the displaced fluid mass) in tested flows. In addition, we will demonstrate that
flow reduced velocity and a dimensionless parameter $c^*$ (defined later in \S \ref{Sec3.1})
are the underlying key parameters responsible for the global optimal
energy harvesting efficiency.

The rest of the paper is structured as follows. The methodology is
introduced in section \S\ref{Sec2}. Then, in section \S\ref{Sec3}, via
ROM, the effects of incoming flow velocity, structural damping, and
mass ratio on the energy harvesting efficiency, are investigated
individually at two Reynolds numbers from the laminar and
subcritical regimes. In addition, the optimal efficiency, damping
ratio, and reduced velocity will be predicted along with
high-resolution maps of the harvesting efficiency. Moreover, the main
results obtained by our ROM will be validated against the experiments
and direct simulations. Finally, discussions and conclusions are drawn in section
\S\ref{Sec4}.

\section{Methodology} 
\label{Sec2}
\subsection{Reduced-order model} \label{Sec2.1}

\begin{figure}[!htbp]
	\centering
	\includegraphics[width=7cm]{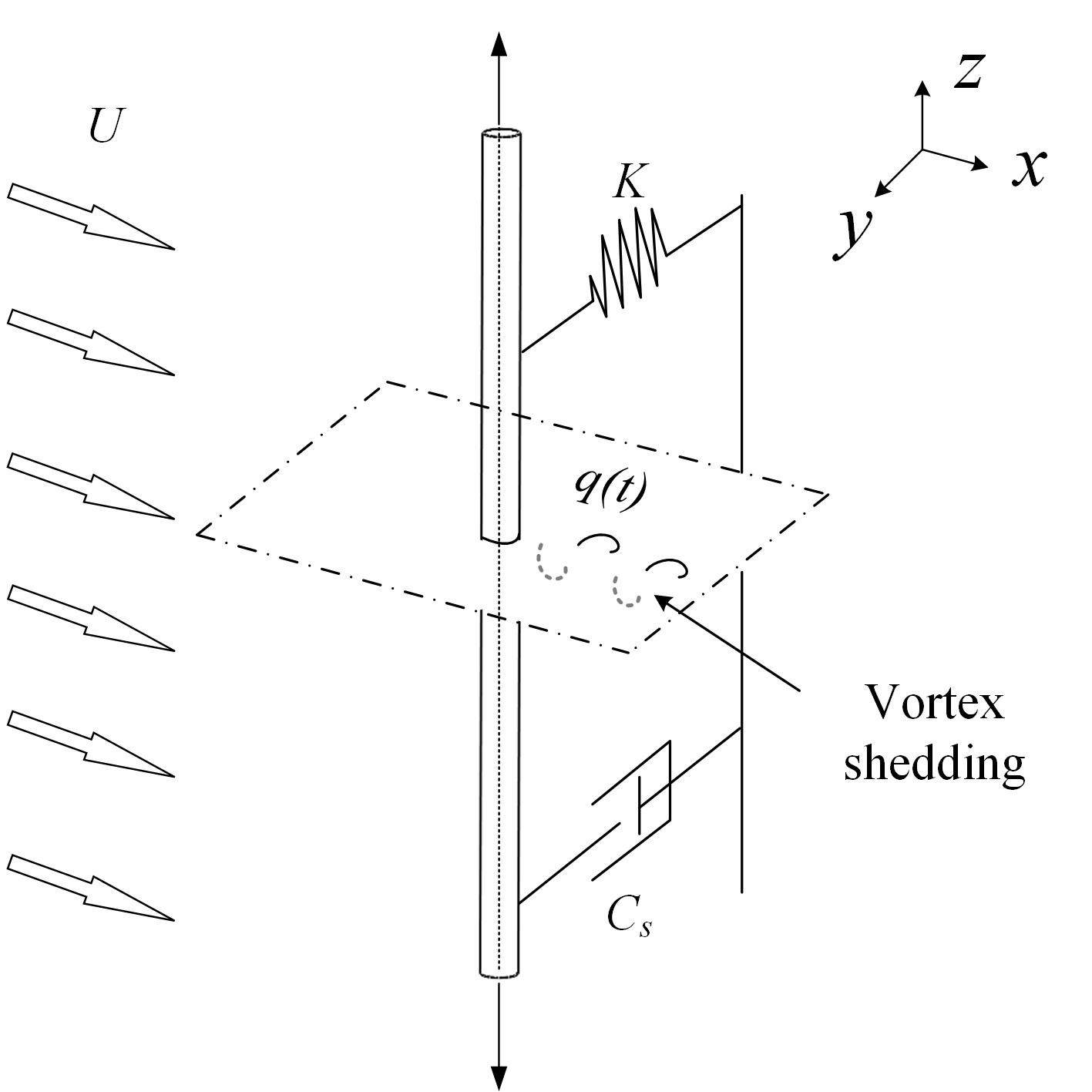}
	\caption{A schematic of the studied problem: an
		elastically mounted circular cylinder subjected to a
		free-stream flow, with a representation of the wake
		oscillator variable $q(t)$. }
	\label{An elastically}
\end{figure}

Figure~\ref{An elastically} shows a schematic for the fluid-structure
system studied: an elastically mounted rigid circular cylinder is
subjected to a free-stream flow, where key parameters are illustrated.
With the external fluid force (i.e. the vortex-induced force $F_v$)
exerted by the vortex shedding, the structure system can be modelled
by a linear second-order mass-spring-damper oscillator, and the
governing equation of motion for the solid cylinder is given by
\begin{equation}
	M\ddot Y + C\dot Y + KY = {F_v},
	\label{eq2.1}
\end{equation}
where $Y$, $\dot{Y}$ and $\ddot{Y}$ represent the displacement,
velocity, and acceleration, respectively, of the vibrating cylinder.
Herein, $M$ represents the total mass including the structural mass
$M_{s}$ and the added mass $M_{a}$. For a circular cylinder, the added
mass can be calculated by $M_{a}=C_{M}\rho D^2 \pi/4$, where $C_{M}$
is a coefficient that can be derived from experiments, see $\S$\ref{Sec2.3}, $\rho$ is the fluid density, and $D$ is
the diameter of the cylinder. $K$ represents the spring stiffness; $C$
is the damping factor, and  it can be computed by $C=C_{s}+C_{f}$, with $C_{s}$
being the structural damping and $C_{f}$ being the added damping
caused by the fluid loading:
\begin{equation}
	{C_f} = \gamma \omega_f \rho D^{2},
	\label{eq2.2}
\end{equation}
where $\gamma$ is a parameter depending on the amplified drag
coefficient $C_D$ \cite{Blevins2001,Facchinetti2004}:
\begin{equation}
	\gamma  = {C_D}/(4\pi {S_t}).
	\label{eq2.3}
\end{equation}
Note that, the above $M$, $K$, $C$ are defined per unit length. In Eq. \ref{eq2.2}, $\omega_f$ is the vortex-shedding angular frequency defined by $\omega_f=2\pi StU/D$, where $St$ is the Strouhal number for flow over a stationary cylinder, and $U$ is the incoming flow velocity.

\citet{Facchinetti2004} proposed a wake oscillator for predicting
the vortex-induced lift $F_{v}$ as follows
\begin{equation}%
	{F_v} = \rho {U^2}DC_L^v/2 = \rho {U^2}Dq{C_{L0}}/4,
	\label{eq2.4}
\end{equation}
where $C_{L0}$ represents the unsteady lift coefficient of a fixed
cylinder. The parameter $q/2$ can be interpreted as a reduced vortex
(or “fluctuating”) lift coefficient that represents the ratio between
the unsteady vortex-induced lift coefficient of an oscillating
cylinder $C_{L}^{v}$ and that of the fixed case ($C_{L0}$)
\cite{Facchinetti2004,Han2021JFM,HanND}. The dimensionless wake
variable $q(T)$, therefore, directly determines the unsteady lift
coefficient caused by vortex shedding, which can be modelled by a van
der Pol non-linear wake oscillator equation, following
\citet[]{Facchinetti2004}:
\begin{equation}
	\ddot q + \varepsilon {\omega _f}\left( {{q^2} - 1} \right)\dot q + \omega _f^2q = f(Y), \quad f(Y)=(A/D)\ddot{Y}.
	\label{eq2.5}
\end{equation}
In this equation, $\varepsilon = 0.3$ and $A=12$ are constant
coefficients derived from experimental data of forced vibrations, see \citet{Facchinetti2004,HanND}. The left-hand side of the
equation is a van der Pol equation with a reference frequency
$\omega_f$ and a growth rate $\varepsilon$, while the right-hand term,
$f(Y)$, is a coupling function of the acceleration $\ddot{Y}$,
connecting the structural dynamics and the wake variable. Now, we
introduce some dimensionless terms as follows:
\begin{equation}
	\begin{array}{l}
		t = T{\omega _f},\quad y = Y/D,\quad {m^*} = 4{M_s}/\pi \rho {D^2},\quad \\
		\quad {U_r} = 2\pi U/({\omega _s}D),\quad \zeta  = {C_s}/(2M{\omega _s}),
	\end{array}
	\label{eq2.6}
\end{equation}
where $t$, $y$, $m^{*}$, $U_{r}$, and $\zeta$ represent
the dimensionless time, dimensionless amplitude, structural mass
ratio, reduced velocity, and damping
ratio, respectively. Here, $\omega_s$ represents the structural angular frequency, defined by $\omega_s=\sqrt {K/M}$. Note that, we use the total mass $M$ instead of the structural mass $M_{s}$ to define the frequency $\omega_s$ and
damping ratio $\zeta$. Substituting equation \eqref{eq2.6} into
equations \eqref{eq2.1}--\eqref{eq2.5} yields the coupled equations
governing the dynamics of the displacement $y(t)$ and the wake
variable $q(t)$
\begin{align}
	\ddot y + (\frac{{2\zeta }}{{{U_r}{S_t}}} + \frac{{4\gamma }}{{\pi {m^*} + \pi {C_M}}})\dot y + \frac{1}{{{U_r}^2{S_t}^2}}y & = \frac{{{C_{L0}}}}{{4{\pi ^3}{S_t}^2({m^*} + {C_M})}} \cdot q, \label{eq2.7}\\[0.5em]
	\ddot q + \varepsilon \left( {{q^2} - 1} \right)\dot q + q &= {A}\ddot y.\label{eq2.8}
\end{align}

In this paper, the two equations
[Eqs. (\ref{eq2.7})(\ref{eq2.8})] in ROM are numerically solved using
a second-order finite difference scheme in time simultaneously. A
small random perturbation on the flow available $q(t)$ is assumed as
the initial condition for the numerical method. The simplicity of the
reduced-order model, i.e., equations \eqref{eq2.7} and \eqref{eq2.8},
allows to predict VIV response at extremely low costs. At present,
this reduced-order model or its slightly modified forms have been
widely and successfully used to predict vortex-induced vibrations
under different conditions (e.g., different $Re$, $m^*$,
$\zeta$, and structural geometry)
with qualitative and quantitative agreements
\citep[e.g.][]{Facchinetti2004,HanND,
	Violette2007,Violette2010,Zanganeh2016,Srinil2012}.

\subsection{Direct Simulation} \label{Sec2.2}

As mentioned previously in the introduction, computations for the
energy harvesting efficiency from VIV at two Reynolds numbers,
$Re = 150$ (in laminar regime) and $Re = 6,000$ (in subcritical
regime), were performed. To further validate our main results
obtained by the proposed reduced-order model, we compared the results
with experiments. However, due to the limitations of our experimental
setup, it is difficult to conduct VIV tests in ultra low-$Re$
laminar flows, and, to the best of the authors' knowledge, so far
there have been no existed published experimental data on energy
harvesting in laminar flows. For this reason, we conducted quasi direct
numerical simulations on VIV to supplement the validations. The direct simulations
was performed by solving the 2D incompressible Navier–Stokes equations
for the fluid field and coupling with a mass-spring-damper oscillator
similar to equation \eqref{eq2.1}.


The governing equations for the fluid flow, including mass and momentum conservation, can be written as 
\begin{align}
	\nabla \cdot \textbf{u} & =0 \label{eq2.9}\\[0.5em]
	\dfrac{\partial \textbf{u}}{\partial T}+(\textbf{u} \cdot \nabla)\textbf{u} &= -\dfrac{1}{\rho}\nabla p+\nu\nabla^2\textbf{u}, \label{eq2.10}
\end{align}
where $\textbf{u}$, $p$, and $\nu$ represent the velocity field,
pressure, and viscosity, respectively. The above equations for the
fluid flow are numerically solved by the segregated pressure implicit
with splitting of operators algorithm. In addition, we use an implicit
first-order scheme for time due to its unconditional stability and a
second-order scheme for both the diffusion and convection terms. The
structural dynamics coupling with the flow field are solved by the
second-order Newmark-beta method, of which the discrete displacement,
velocity, and acceleration of the vibration can be respectively
written by
\begin{align}
	\ddot{y}_{t+\Delta{t}} &= \dfrac{1}{\beta_{n} \Delta{t}} \left[ (y_{{t}+\Delta{t}}-y_{{t}})\dfrac{1}{\Delta{t}}-\dot{y}_{{t}}\right]-\left(\dfrac{1}{2\beta_{n}}-1\right)\ddot{y}_{{t}} \label{eq2.11}\\[0.5em]
	\dot{y}_{{t}+\Delta{t}} &= \dot{y}_{{t}}+\left[\left(1-\gamma_{n} \right)\ddot{y}_{{t}}+ \gamma_{n} \ddot{y}_{{t}+\Delta{t}}\right]\Delta{t}\label{eq2.12}\\[0.5em]
	{y}_{{t}+\Delta{t}} &= {y}_{{t}}+ \dot{y}_{{t}}\Delta{t}+ \left[\left(\dfrac{1}{2}-\beta_{n}\right)\ddot{y}_{{t}}+ \beta\ddot{y}_{{t}+\Delta{t}}\right]\Delta{t}^2, \label{eq2.13}
\end{align}
where $\beta_{n}=0.25$ and $\gamma_{n}=0.5$ are two constants. More
details of the above FSI coupling approaches can be found in
the previous studies
\cite{HanAIP,Han2021JFM,HanOE,HanPOF}.

\begin{figure}
	\centering
	\includegraphics[width=12.5cm]{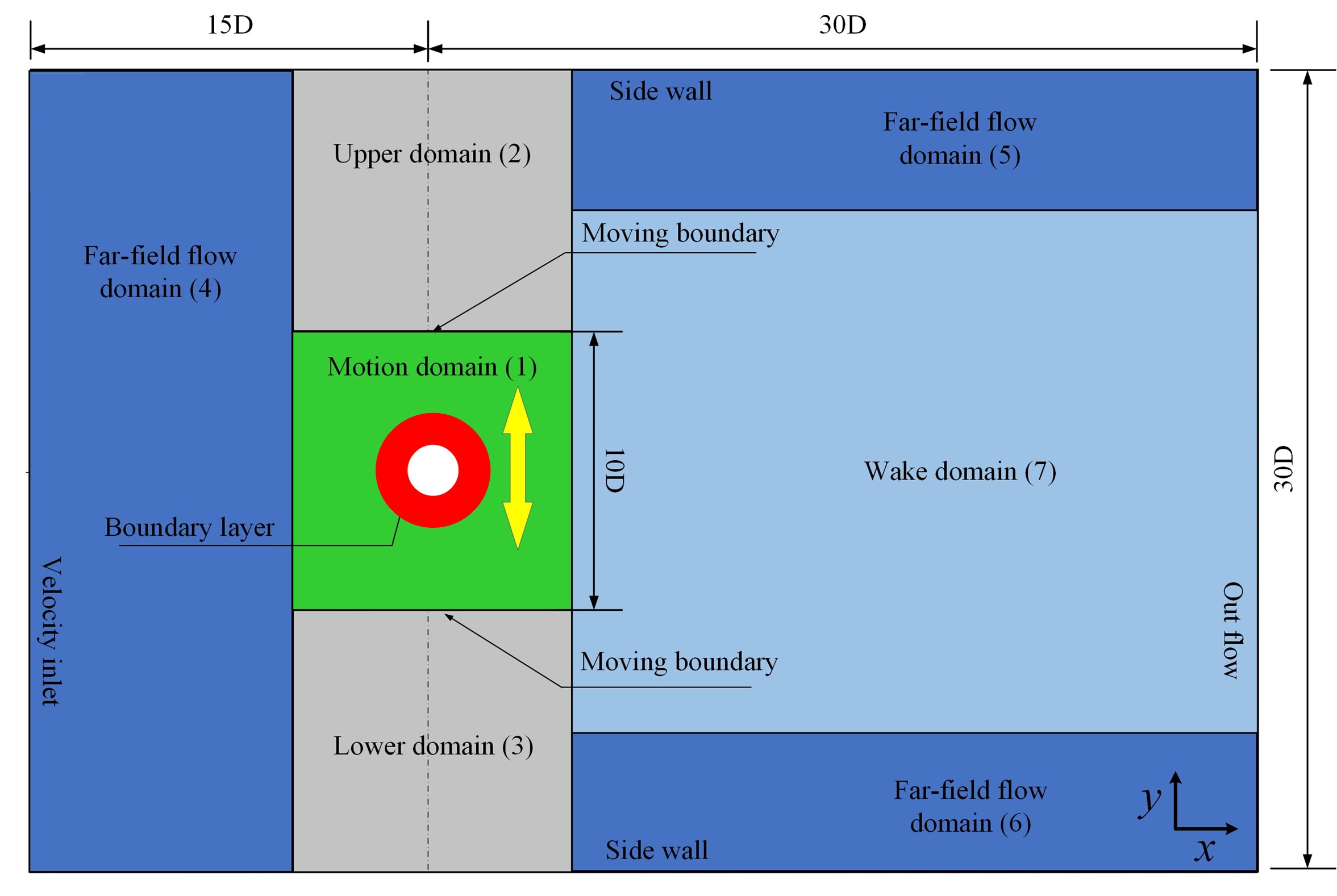}
	\caption{Schematic of the computational domain.}
	\label{Schematic of}
\end{figure}
Figure~\ref{Schematic of} shows the computational domain used in the
numerical simulations, which is of 45D in length and 30D in width. We
split the entire domain into seven sub-regions, in order to apply the
block dynamic mesh technique. The grids in motion domain (1) are dense
and are not updated during the vibration to ensure the numerical
accuracy. For more details of the mesh strategy, initial conditions,
and boundary conditions, see \cite{HanAIP,HanOE,Wang2020}.

\subsubsection{Validations} \label{Sec2.2.2}
\begin{figure}
	\centering
	\includegraphics[width=10cm]{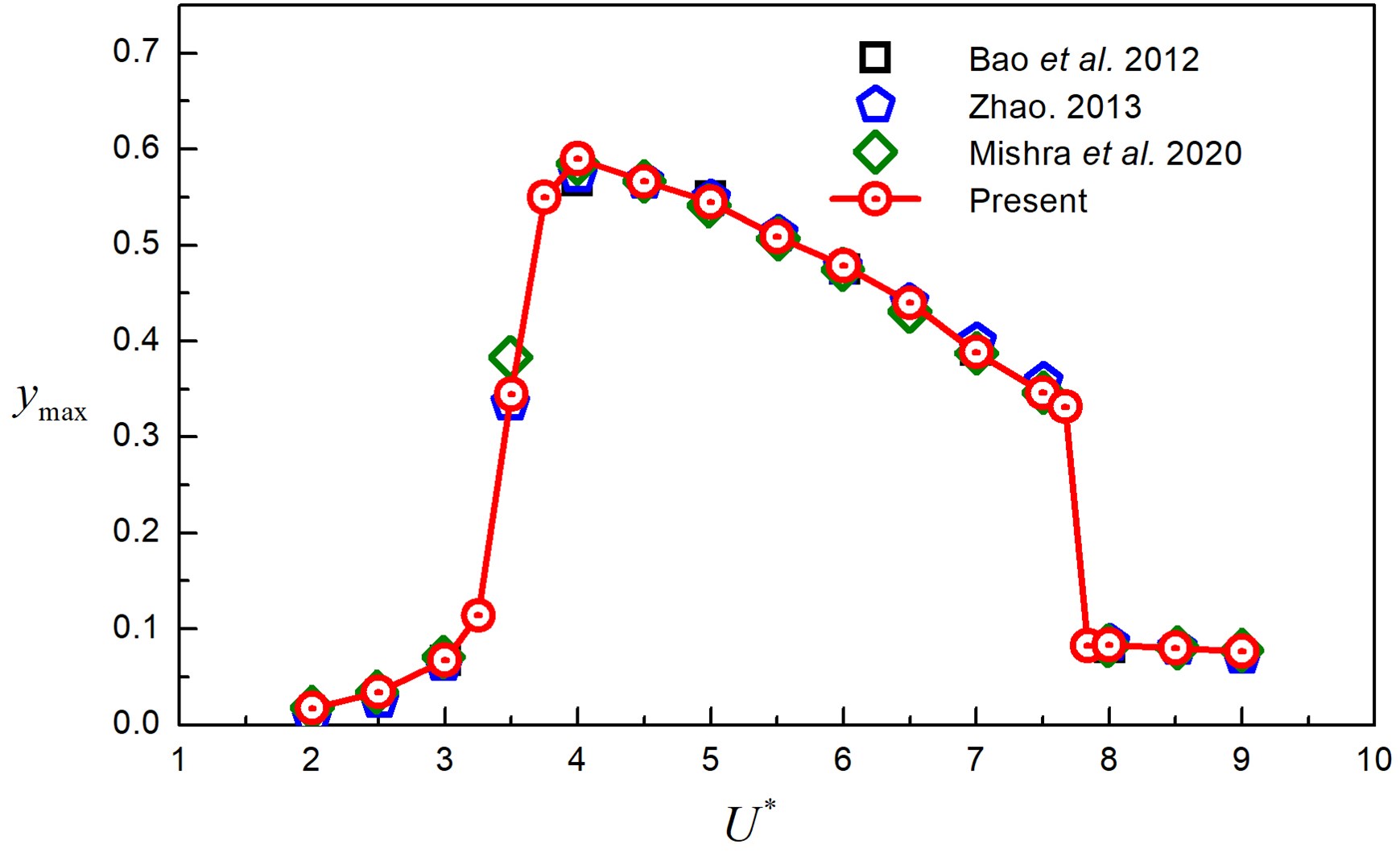}
	\caption{Comparison between the DNS-FSI solver used in the
		present paper and the previous numerical studies of
		\citet{Bao2012,Mishra2020,Zhao2013}. Parametric settings:
		$Re = 150$; $m^*=2.546$; $\zeta=0$. Here, $y_{max}$ presents
		the maximum normalised amplitude.}
	\label{Comparison for DNS-FSI}
\end{figure}

So far, the quasi-DNS-based fluid-structure interaction (refereed to as
DNS-FSI) solver for VIV of a circular cylinder has been developed. In
this section, we validate it against published data. \citet{Bao2012},
\citet{Mishra2020}, and \citet{Zhao2013} conducted numerical studies of VIV
of a circular cylinder with low mass and damping ratios $m^* = 2.546$
and $\zeta = 0$ at $Re=150$. Using the same parametric settings of
$m^*$, $\zeta$, and $Re$, we applied our DNS-FSI solver to predict the
VIV amplitude response and compared with the previous studies in
figure~\ref{Comparison for DNS-FSI}. It should be noted that the total
grid nodes and dimensionless time step for the present DNS-FSI solver
were reasonably set to $10^{5}$ and $\Delta T=0.002s$
$(\Delta t=0.01)$, considering the results of mesh and time step
independence tests. To be consistent with the compared studies, we
define the reduced velocity by $U^{*}=2\pi U/(D\sqrt{K/M_{s}})$,
without considering the effect of added mass. The validation results
in figure~\ref{Comparison for DNS-FSI} show excellent agreements among
the present DNS-FSI solver and the previous DNS studies. In addition,
we predict the hydrodynamic characteristics of flow over a fixed
circular cylinder at $Re=150$ for further validation, using the
proposed DNS-FSI approach without solving the structural dynamics.
Table~\ref{table1} presents the results of the root-mean-square
(r.m.s.) of the lift coefficient $C_{L0}$, the time-averaged drag
coefficient $C_{d}$, and the Strouhal number $St$, in comparison with
other published studies, confirming again the accuracy of our DNS-FSI
solver.

\renewcommand\arraystretch{1.25}
\begin{table}[htb]\footnotesize
	\centering
	\newcommand{\topcaption}{%
		\setlength{\abovecaptionskip}{0pt}%
		\setlength{\belowcaptionskip}{10pt}%
		\caption} \topcaption{A comparison between the preset
		results and previous published data, of flow over a
		fixed circular cylinder at $Re=150$.}
	\setlength{\tabcolsep}{4.5mm}{
		\begin{tabular}{cccc}
			\toprule  
			Literature & $St$& Time-averaged $C_{d}$ & r.m.s. of $C_{L0}$\\
			\hline
			\cite{Norberg2003}  & 0.183 & \rule[4pt]{8pt}{0.5pt} & 0.356 \\
			\cite{Qu2013}   & 0.184 & 1.305 & 0.355\\
			Present  & 0.185 & 1.331 & 0.358 \\
			\hline
			\hline
		\end{tabular}
	}
	\label{table1}
\end{table}

\subsection{Experimental details} \label{Sec2.3}%
The present experiments were conducted in a recirculating
free-surface water channel of the {\em Fluids Laboratory for
	Aeronautical and Industrial Research (FLAIR)} at Monash
University. This water channel has a test section of $600$\,mm
in width, $800$\,mm in depth and $4000$\,mm in length, and the
water flow velocity can be varied in the range of $45$\,mm/s
$\leqslant U \leqslant $ 450\,mm/s, with the turbulence level
less than $1\%$. The mass-spring-damper oscillator was modelled
on a low-friction air-bearing system. More details and
validation studies of the experimental methodologies used can be
found in the previous related studies of
\citet{zhao2018Dsection,SotiJFS2018,wong2018}.

\begin{figure}
	\centering
	\includegraphics[width=11cm]{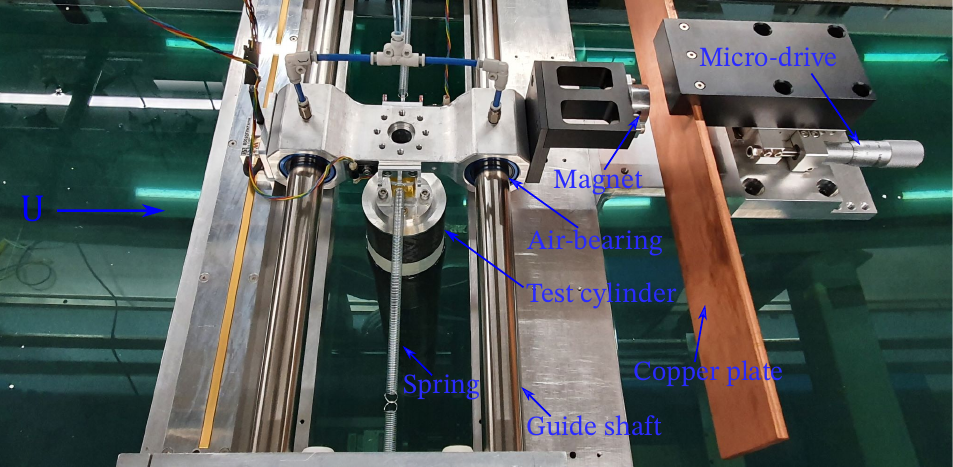}
	\caption{A photograph showing the experimental set-up and its
		key components.}
	\label{exp_photo}
\end{figure}

The test rigid circular cylinder was precision-made to have an
outer diameter of $D = 30\pm0.01$\,mm. The immersed length of
the cylinder was $L = 614$\,mm, giving a span-to-diameter ratio (aspect ratio)
of 20.5. The displaced mass of fluid (water) was
$M_d = \rho \pi D^2L/4 = 433.6$\,g, while the total oscillating
mass was $M_s = 2630.6$\,g, yielding a mass ratio of
$m^* = M_s/M_d = 6.07$.

\begin{figure}
	\centering
	\includegraphics[]{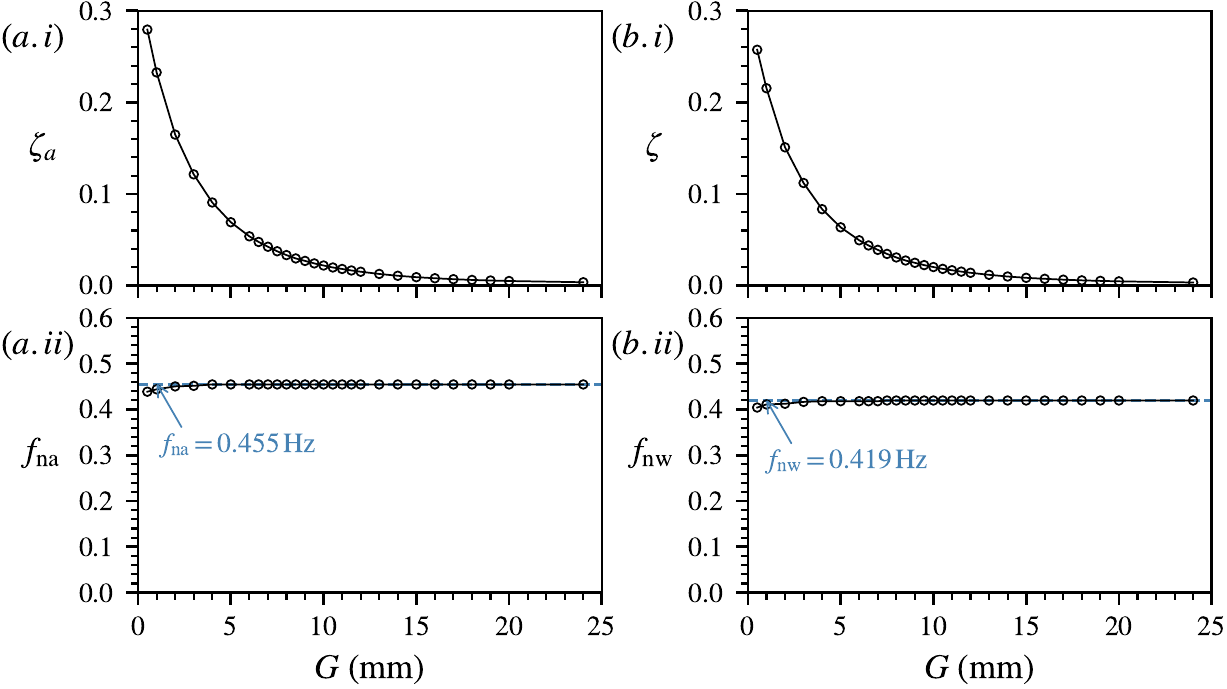}
	\caption{Free decay test results of the structural
		damping ratios and natural frequencies as a function of
		the damper gap. Panels (a.i) and (a.ii) show the results
		of the structural damping ratio ($\zeta_a$) and the
		natural frequency ($f_{na}$) in air, respectively, while
		panels (b.i) and (b.ii) show the structural damping ratio
		with the consideration of added mass ($\zeta$) and the
		natural frequency ($f_{nw}$) in quiescent water,
		respectively.}
	\label{fig:damping_gap}
\end{figure}

To control the structural damping ratio, an eddy-current-based
damping device was employed, which consisted of a micro-drive
stage with a resolution of $0.01$\,mm to vary the damper gap for
the structural damping control. Photographs of this damping
device can be found in the studies of
\citet{SotiJFS2018,zhao2022prf}. The structural stiffness of the
air-bearing system was given by precision extension springs.
Free decay tests were performed to measure the natural
frequencies of the mass-spring-damper system:
$f_{na} = 0.455$\,Hz in air and $f_{nw} = 0.419$\,Hz in still
waters, while the structural damping ratio with consideration of
added mass was given by $\zeta = C/2\sqrt{(K(M_s + M_a))}$, with
$M_a = ((f_{na}/f_{nw})^2 - 1)M_s = 471.3$\,g and
$C_M = M_a / M_d = 1.08$. Figure~\ref{fig:damping_gap} shows the results of the
structural damping ratios and natural frequencies from free
decay tests in both quiescent air and water
\cite{zhao2022prf}.

\section{Results} \label{Sec3}%
In this section, we present four high-resolution maps of the
energy harvesting efficiency as a function of reduced velocity
and damping ratio for different mass ratios and Reynolds
numbers. These maps are obtained from our ROM results.
Considering the definition of mass ratio in
equation~\eqref{eq2.6}, for a given VIV energy harvester system,
the structural mass $M_{s}$ is fixed, while the greater the
density of the fluid medium $\rho$ the lower the mass ratio
$m^{*}$ is. Thus, two (low and high) mass ratios, $m^{*}=6$ and
$500$, are used to represent as the ocean and wind VIV energy
converters, respectively, given the different fluid densities
between water and air. Two Reynolds numbers are tested in the
present work: the $Re=150$ case is in the typical
laminar regime, while the $Re = 6,000$ case is in the
subcritical regime.

Considering one vibration cycle period $T_{vib}$, the harnessed power
by VIV can be computed by
\begin{equation}\label{eq2.14}
	{P_h} = \frac{1}{{{T_{vib}}}} \int_{0}^{T_{vib}} {F_{tot}}\frac{{dY}}{{dT}}dT = \frac{1}{{{T_{vib}}}}  \int_{0}^{T_{vib}} ({M_s}\frac{{{d^2}Y}}{{d{T^2}}} + {C_s}\frac{{dY}}{{dT}} + KY)\frac{{dY}}{{dT}}dT,
\end{equation}
where $F_{tot}$ is the total force acting on the vibrating cylinder. 
Note that the total force can be decomposed by $F_{tot} = F_v – F_a - C_f\dot{Y}$, with $F_a$ is the added-mass forces.
In fact, one can find that only the velocity term $\dot{Y}$ can
contribute to the harnessed power, while the other terms that related
to the $\ddot{Y}$ and $Y$ will become zero after integration
\citep[see][]{SotiJFS2018, HanOE}. Thus, equation~\eqref{eq2.14} can
be rewritten by
\begin{equation}\label{eq2.15}
	{P_h} = \int_{0}^{T_{vib}}\frac{1}{{{T_{vib}}}}({C_s}\frac{{dY}}{{dT}})\frac{{dY}}{{dT}}dT = \left\langle {{C_s}{{(\frac{{dY}}{{dT}})}^2}} \right\rangle,
\end{equation}
where $<\cdot>$ represents the time-averaged operation. To estimate
the energy harvesting efficiency, we introduce the power produced by
the fluid flow:
\begin{equation}\label{eq2.16}
	{P_f} = \frac{1}{2}\rho {U^3}(D),
\end{equation}
Note that, there are in fact several criteria to define ${P_f}$, depending on using $D$ or $D+2y_{max}$, as the projected area (see for instance \cite{Bernitsas2008,Grouthier2014,Ding2015}). A different definition of $P_f$ will not affect the results, while it is only related to the data post-processing. Here, we use the same definition as in Refs \cite{Ding2015,Grouthier2014} and our previous studies \cite{SotiJFS2018,HanAIP,zhao2022prf} for consistency. Now, the efficiency $\eta$ can be defined as the ratio of the harnessed power to the fluid power available, namely
$\eta = P_h / P_f$. By substituting the dimensionless time $t$,
damping ratio $\zeta$, dimensionless amplitude $y$, mass ratio $m^*$,
and reduced velocity $U_r$ into the equation for $\eta$, it gives
\begin{equation}\label{eq2.17}
	\eta  = \frac{{8{\pi ^4}{St}^2({m^*} + {C_M})\left\langle {\zeta \dot y\dot y} \right\rangle }}{{{U_r}}}.
\end{equation}

\subsection{Energy harvesting from VIV at high-$Re$  in the subcritical regime} \label{Sec3.1}

\begin{figure}
	\centering
	\includegraphics[width=15.8cm]{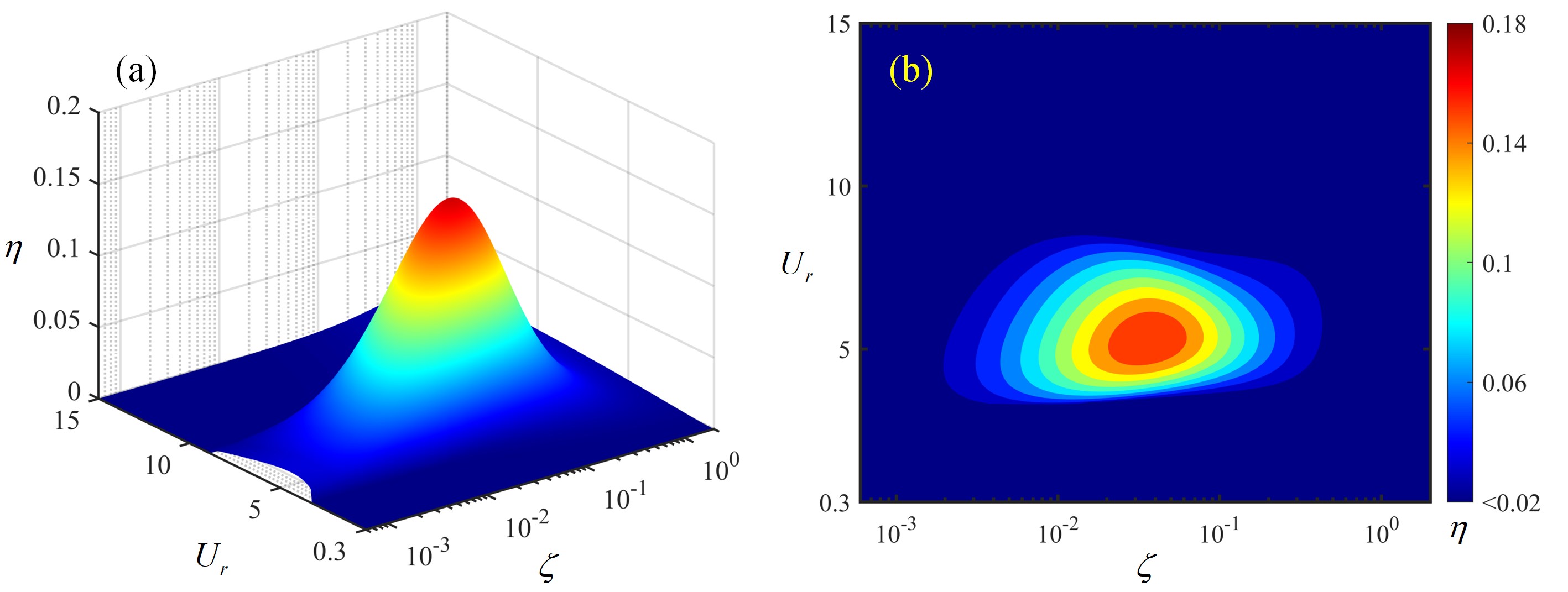}
	\caption{Contour maps of the energy harvesting efficiency
		$\eta$ as a function of the reduced velocity $U_{r}$ and
		damping ratio $\zeta$, at $m^* = 6.07$ and
		$Re \approx 6 \times 10^3$. The right plot shows the contour
		map of the left one in a two-dimensional plane.}
	\label{HighRe_Lowmass}
\end{figure}

As can be seen from equations~\eqref{eq2.15} and \eqref{eq2.17}, the
energy harvesting efficiency $\eta$ is strongly determined by the
damping ratio $\zeta$, reduced velocity $U_{r}$ and mass ratio
$m^{*}$. For a given VIV energy harvester, the structural mass ratio
is generally fixed, while the other two terms may vary. Here, we apply
the reduced-order model to optimise the efficiency as a function of
$U_r$ and $\zeta$, individually at a high mass ratio $m^{*} = 500$
(for wind VIV energy converter) and a low mass ratio $m^{*} = 6$
(for ocean energy converter). The tested Reynolds number is around
$6\times10^{3}$. The flow parameters, i.e., the lift coefficient
$C_{L0}=0.7$ and Strouhal number $St=0.208$ of flow over a fixed
cylinder, used for our ROM are taken from experiments under the same
Reynolds number \cite{Norberg2003,zhao2018inline}. Figure \ref{HighRe_Lowmass}
shows the full efficiency map spanning the reduced velocity range
$0.3 \leqslant U_r \leqslant 15$ and the damping ratio range
$0.0001\leqslant \zeta \leqslant 10$, at $m^{*} = 6$. Both the
ranges of $U_{r}$ and $\zeta$ are wide enough to cover the
high-efficiency region, while the efficiency outside the computed
space is negligible. The results in figure~ \ref{HighRe_Lowmass}
indicate that $\eta$ first increases and then experiences a decrease
with increasing $U_{r}$. A comparable scenario can be found for the
effects of damping ratio on the efficiency. Overall, the maximum
efficiency is found to be $\eta\approx0.17$, while, correspondingly, the
optimal reduced velocity and damping ratio are $U_{r}=5.275$ and
$\zeta=0.0379$, respectively. One can find that the optimal $U_{r}$ is
slightly higher than the resonance velocity $1/St$, indicating the maximum
efficiency locates among the lock-in range, consistent with the
previous studies of Refs\cite{Bernitsas2008,SotiJFS2018}.

\begin{figure}
	\centering
	\includegraphics[width=12.5cm]{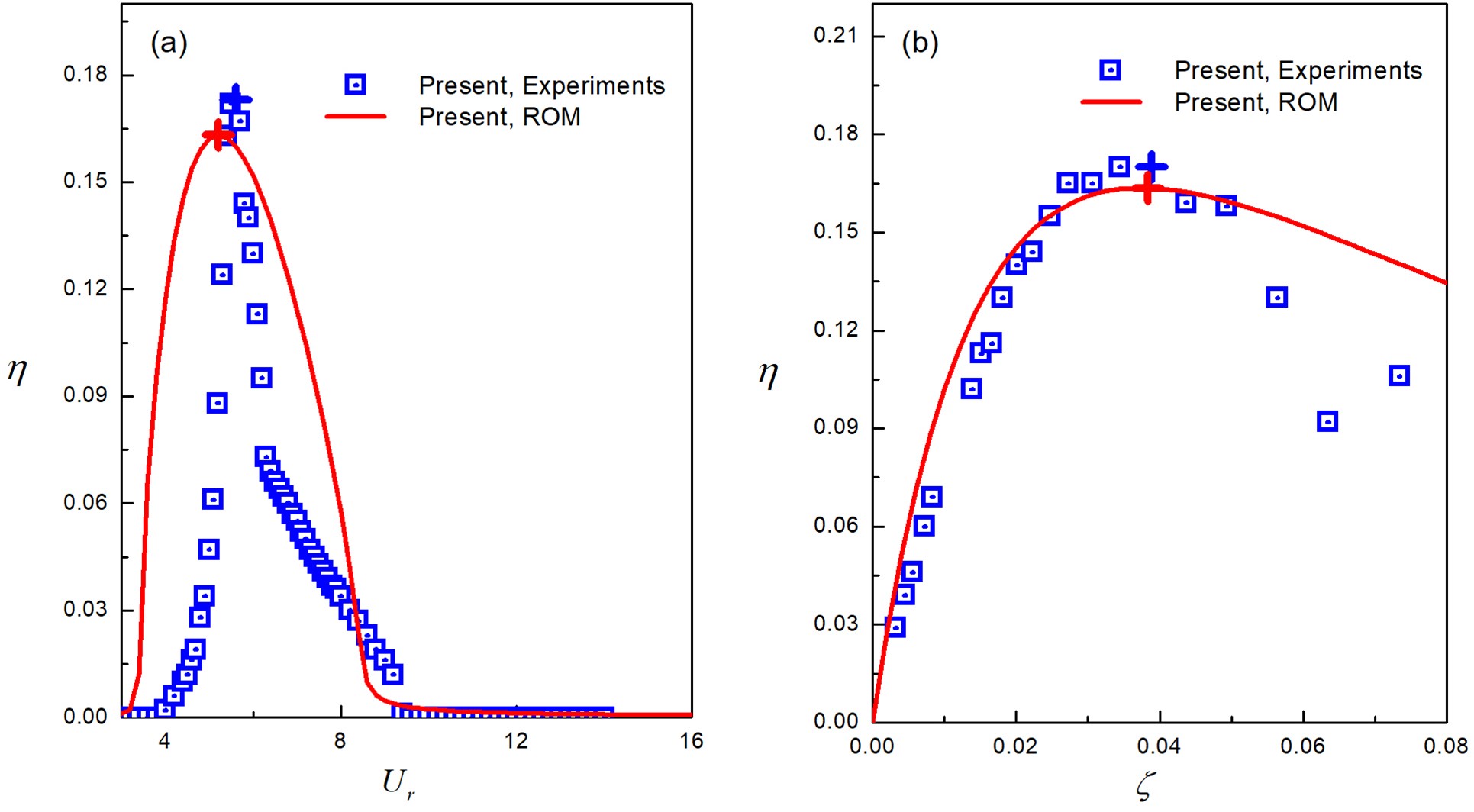}
	\caption{Comparisons of the variations of energy harvesting
		efficiency $\eta$ as a function of the $(a)$ reduced
		velocity $U_{r}$ and $(b)$ damping ratio $\zeta$, obtained
		by our ROM and experiments. "$+$" presents the maximum
		efficiency point.}
	\label{HighRe_Validation}
\end{figure}

To further validate our predicted results obtained by the ROM, we
conducted experiments on VIV of a circular cylinder under the same
flow conditions and structural parameters as numerical predictions.
First, to validate the ROM results for the effect of velocity $U_{r}$,
the damping ratio was kept unchanged at $\zeta=0.0305$ near the
optimal value, while the reduced velocity $U_{r}$ was varied from 2.8
to 14. Figure~\ref{HighRe_Validation}$(a)$ shows a comparison of
$\eta$ obtained from our experiments and ROM. It can be seen that our
ROM can capture the maximum efficiency, the resonance phenomenon, and
most features due to the influence of reduced velocity. More
specifically, the value of maximum efficiency predicted by ROM is
0.164, observed at $\zeta=0.0305$, which is highly close to the
experimental value $0.173$. Also, the optimal reduced velocity in
figure \ref{HighRe_Validation}$(a)$ obtained by our ROM and
experiments are similar, i.e., $U_{r}=5.3$ for ROM and $U_{r}=5.6$ for
experiments. Second, keeping the reduced velocity $U_{r}$ fixed at 5.5
near the optimal value, we compare the effect of damping ratio
predicted by the ROM and that obtained from our experiments in figure
\ref{HighRe_Validation}$(b)$. The experimental results show the
maximum efficiency 0.173 occurring at $\zeta=0.03445$ and $0.0388$,
while the ROM sees the predicted maximum efficiency $\eta=0.164$ and
the optimal damping ratio $\zeta=0.0383$ in excellent agreement with
the experiments. However, some differences can still be found between
our ROM and experiments. This is mainly because the body oscillation
is not perfectly in phase with the acceleration, particularly for a
low-mass-ratio body in the VIV lower branch
\citep[see][]{Williamson2004,deLangre2006}. Thus, the proposed ROM,
coupling with the body acceleration $\ddot{y}$ [see
equation~\eqref{eq2.8}], cannot capture the classical lower branch of
VIV \cite{Facchinetti2004}, which leads to some discrepancies on the
phase angle prediction. Perhaps, introducing an out-of-phase term
(i.e., the velocity $\dot{y}$) coupled with the acceleration term
$\ddot{y}$ in equation~\eqref{eq2.8} may improve the accuracy for the
predictions of phase and thus the amplitude and efficiency; however,
this would make the ROM become more complicated, which is beyond the
scope of present study. In addition, more accurate fits with the
experimental results could probably be obtained if the coefficients
used in our ROM were modified. This would not be illegitimate as, for
instance, the lift coefficient $C_{L0}$ and Strouhal numbers $St$ have
been found to vary over certain ranges -- for more details, see the
review study of \citet{Norberg2003}. A sensitive analysis of the
parameters $C_{L0}$ and $S_{t}$ on the ROM has been reported in the
recent work of \citet{Han2021JFM}. In general, by comparing with
present experiments, we can conclude that the proposed ROM, with
extremely low costs, can qualitatively and to some extent
quantitatively predict the energy harvesting efficiency from the
complex FSI phenomenon -- vortex-induced vibration.

\begin{figure}
	\centering
	\includegraphics[width=15.5cm]{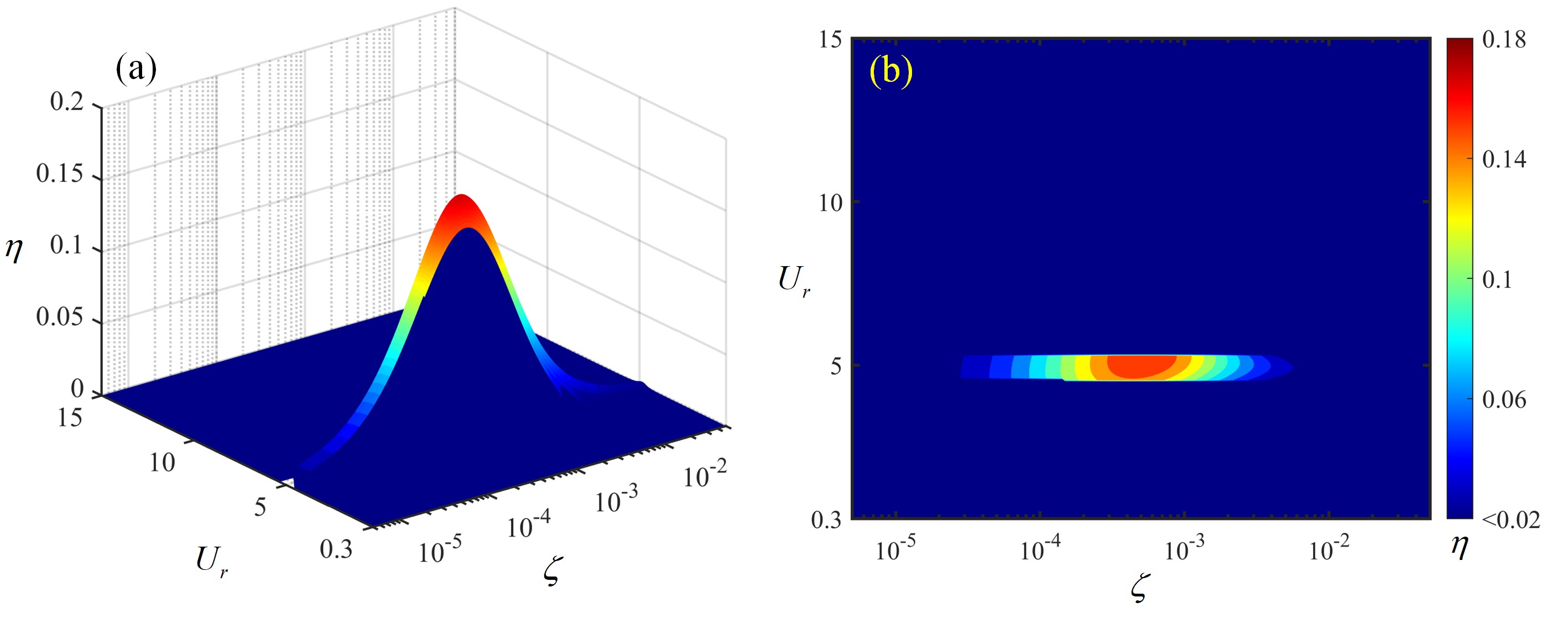}
	\caption{Contour maps of the energy harvesting efficiency
		$\eta$ as a function of the reduced velocity $U_{r}$ and
		damping ratio $\zeta$, at $m^* = 500$ and
		$Re \approx 6\times 10^3 $. The right plot shows the contour
		map of the left one in a two-dimensional plane}
	\label{HighRe_Highmass}
\end{figure}

In terms of the high-mass-ratio oscillator (i.e., to represent the
wind VIV energy harvester), figure \ref{HighRe_Highmass} presents a
full map of the efficiency predicted using the ROM. In this case, the
applied parameters remain the same as in figure~\ref{HighRe_Lowmass},
except $m^{*}=500$. The results show that the peak value of efficiency
$\eta=0.17$ is barely affected by the mass ratio. The optimal reduced
velocity and damping ratio for this mass ratio are found to be $5.25$
and $5.168\times 10^{-4}$, respectively. The optimal reduced
velocities $U_{r}$ for the two mass ratio cases tested are almost the
same, which are close to the resonance velocity. It is interesting to
note that the optimal damping ratio for the low-mass-ratio case in
figure~\ref{HighRe_Lowmass} is approximately $73$ times of that of the
high-mass-ratio case. Moreover, considering the effect of added mass,
the two mass ratio cases tested, i.e., $6 + 1.08$ and $500 + 1.08$, also
differ similarly by a factor of $71$. This seems to suggest that the location
of the maximum efficiency point $\eta$ is controlled by the reduced
velocity and the product of the total mass and damping ratios
$(m^{*}+C_{M})\zeta$. In fact, the coupled mass-damping parameters or its modified forms
(such as Skop-Griffin parameter $S_{G}$ and Scruton number $S_{c}$) are known to determine the
peak amplitude of VIV \cite{Blevins2001, Williamson2004}. However, as reported by \cite{Vandiver2012}, the parameter $(m^{*}+C_{M})\zeta$ also has its limitations. \cite{Vandiver2012} derived a parameter $c^*$, namely $c^*=2C_s\omega_v/\rho U^2$ ($\omega_v$ is the angular frequency of vibrations), to replace the coupled mass-damping parameter to collapse the VIV amplitudes. Following \cite{Vandiver2012}, we here show that the parameter $c^*$ can be useful to control the maximum energy harvesting efficiency point as well. The parameters $c^*$ for the global optimal efficiency are identical for both the two mass ratios: it is 1.14 for $m^*=6$ and 1.16 for $m^*=500$. More details
and analysis on this are given in section \S\ref{Sec4}.


\begin{figure}
	\centering
	\includegraphics[width=7cm]{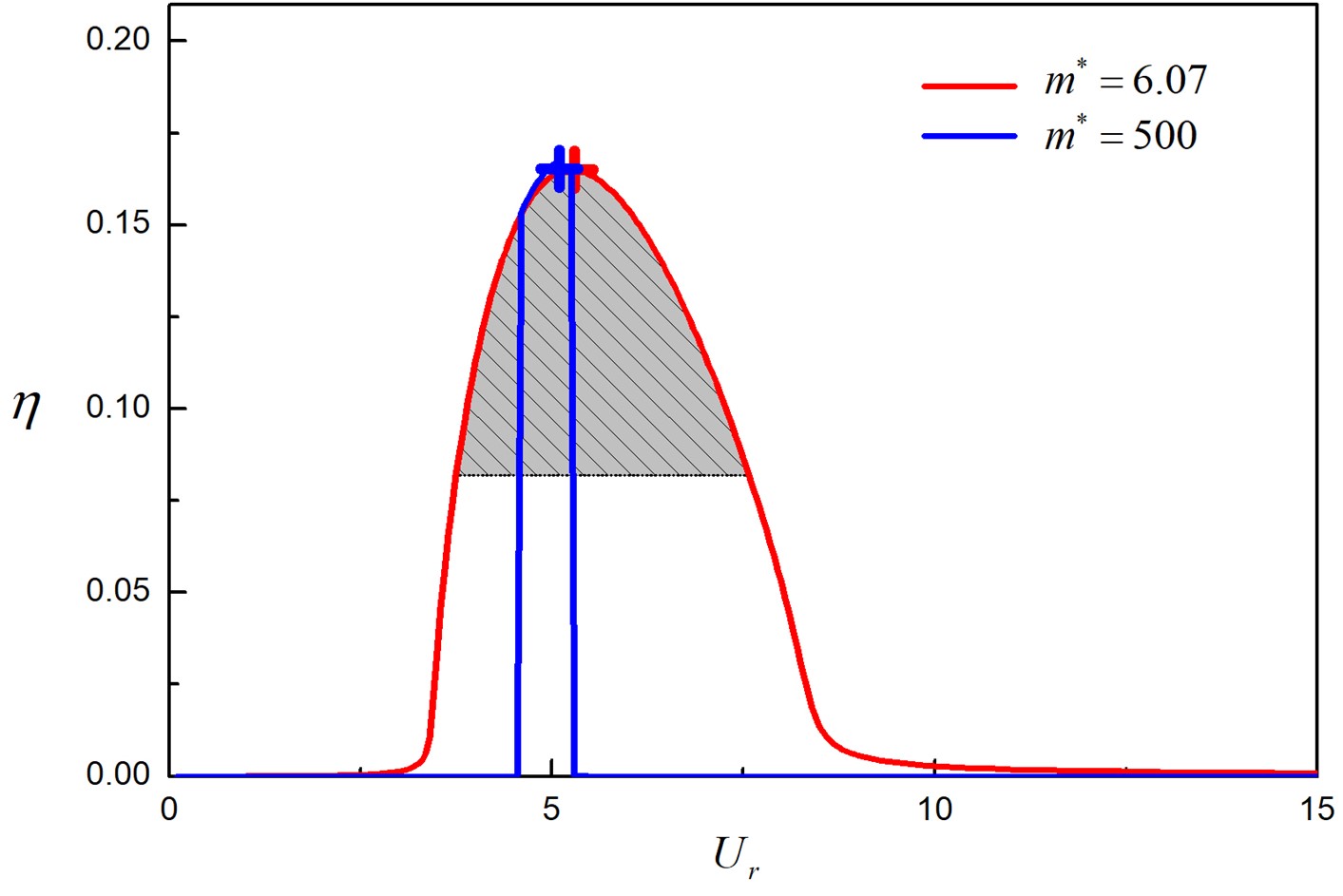}
	\caption{Comparison of the energy harvesting efficiency $\eta$
		as a function of $U_{r}$ at the optimal damping ratio
		between a low and a high mass ratio. The $+$ symbols
		represent the locations of the maximum efficiency for each
		mass ratio case.}
	\label{Comparsion_at_optimaldamping_highRe}
\end{figure}

To better compare the energy harvesting performance at low and high
mass ratios, figure \ref{Comparsion_at_optimaldamping_highRe} shows
the variation of $\eta$ with $U_r$ at the optimal damping ratios for
$m^* = 6$ and $500$. Note that, as mentioned previously in Section
\ref{Sec2.1}, the present paper uses a small random perturbation on
the $q(t)$ as the initial condition to start the ROM computations, assuming the structural body vibrates from rest.
Actually, a different initial condition (small/large/increasing velocity/decreasing velocity perturbations on
the system) will give the different predictions for a high-mass
system, mainly because of inertia effects. A large perturbation or increasing velocity initial conditions on the
FSI system, particularly a high-mass-ratio system, allow to yield a wide range of lock-in, but
the maximum efficiency point is still the same as the predictions with
a small perturbation. The effect of the initial conditions however is
out of the scope of the present paper and will not discussed in
detail. Here, we define the high-efficiency region among which the
efficiency $\eta$ is larger than the half of the maximum value
observed. As can be seen, although the maximum efficiency is almost
identical for the two mass ratios, the high-efficiency regime for the
case of $m^* = 6$ is obviously wider than that of $m^{*}=500$. This
phenomenon is attributable to the fact that the lock-in (or
synchronisation) range, where significant energy harvesting
performance is achieved, for the low-mass-ratio case is much wider
than that of the high-mass-ratio case. Note that a lighter body can
enhance the fluid-structure interaction and thus the wake
instabilities responsible for the cause of the structural vibration
\citep[see][]{deLangre2006}. By comparing the results in
figures~\ref{HighRe_Lowmass}, \ref{HighRe_Highmass}, and
\ref{Comparsion_at_optimaldamping_highRe}, it can be concluded that
the energy harvesting efficiency is more robust for a low-mass-ratio
system, indicating that VIV energy harvesters should perform better in
water than in air.

\subsection{Energy harvesting from VIV in a laminar
	flow} \label{Sec3.2}

\begin{figure}
	\centering
	\includegraphics[width=14.5cm]{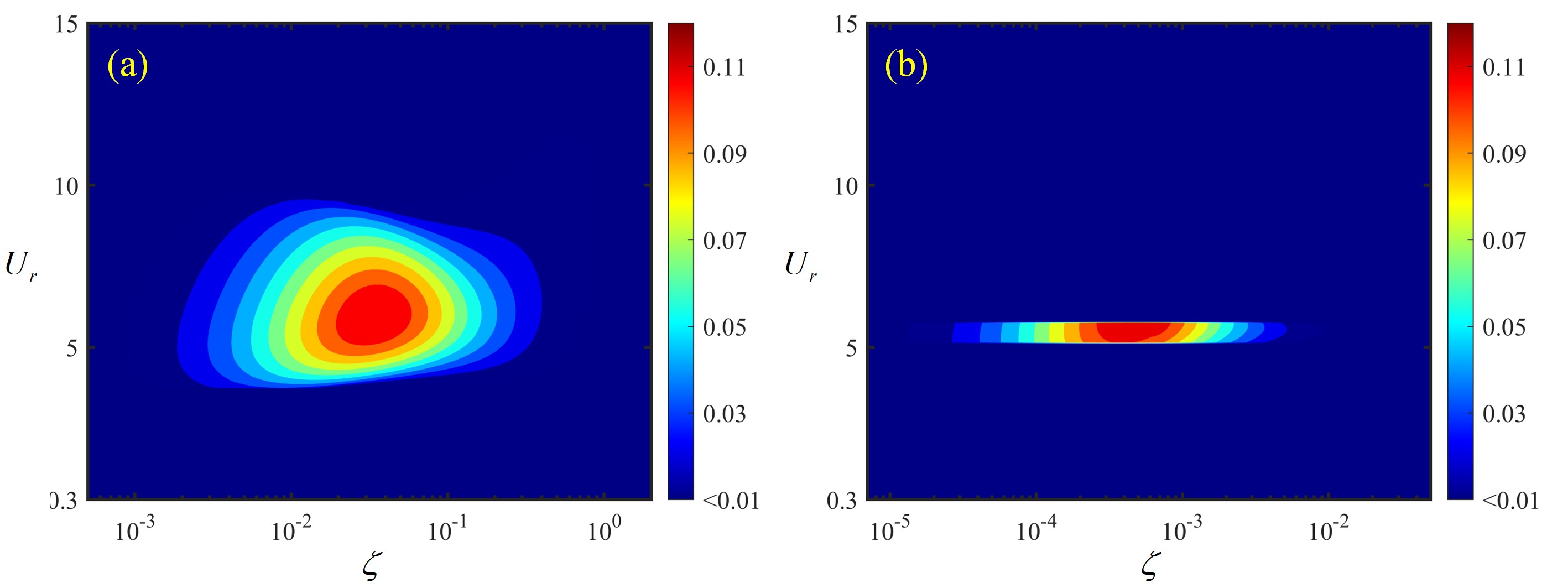}
	\caption{Optimisation of the energy harvesting efficiency
		$\eta$ as a function of $U_{r}$ and $\zeta$ in a low-Re flow at $(a)$
		$m^* = 6.07$ and $(b)$ $m^* = 500$.}
	\label{optimazition for low Re}
\end{figure}

In this section, we investigate the characteristics of the energy
harvesting efficiency from VIV in a laminar flow. Using the parameters
$C_{L0}^{r.m.s.}=0.36$ and $S_{t}=0.185$ taken from table~\ref{table1}
for flow over a fixed circular cylinder at $Re=150$, we construct the
full energy harvesting efficiency maps for the two mass ratios,
$m^* = 6$ and $500$, in a laminar flow. As can be seen in figure~
\ref{optimazition for low Re}, similar effects of $U_{r}$, $\zeta$,
and $m^{*}$ on the efficiency are observed for two mass ratio cases at
$Re = 150$ (figures~\ref{HighRe_Lowmass} and \ref{HighRe_Highmass}).
For $m^* = 6$ in figure~\ref{optimazition for low Re}$(a)$, the
maximum efficiency is observed to be 0.118, occurring at $U_r=5.92$ and
$\zeta=0.0338$. On the other hand, for $m^* = 500$ in
figure~\ref{optimazition for low Re}$(b)$, the maximum efficiency is
found to be $\eta=0.119 $ and the optimal velocity $U_r=5.90$, similar
to the case of $m^* = 6$, while the optimal damping ratio is 70
times lower than that of the case of $m^* = 6$. Note that the product of the optimal damping ratio and the total mass ratio is found to be 0.239 and
0.241 for $m^* = 6$ and $500$, respectively. The
parameter $c^*$ for the heavy ($m^* = 500$) and light ($m^* = 6$)
FSI systems at the optimal efficiency points are 0.83 and 0.84,
respectively. The above two nearly identical values indicate that
the maximum energy harvesting efficiency is predominantly controlled
by $c^*$, regardless of the present tested two
Reynolds numbers. Comprehensively from the results in
figures~\ref{HighRe_Lowmass}, \ref{HighRe_Highmass} and
\ref{optimazition for low Re}, in general, the overall performance of
energy harvesting from VIV at $Re = 6,000$ from the subcritical
regime is better than that at $Re = 150$ from the laminar regime.



\begin{figure}
	\centering
	\includegraphics[width=12.5cm]{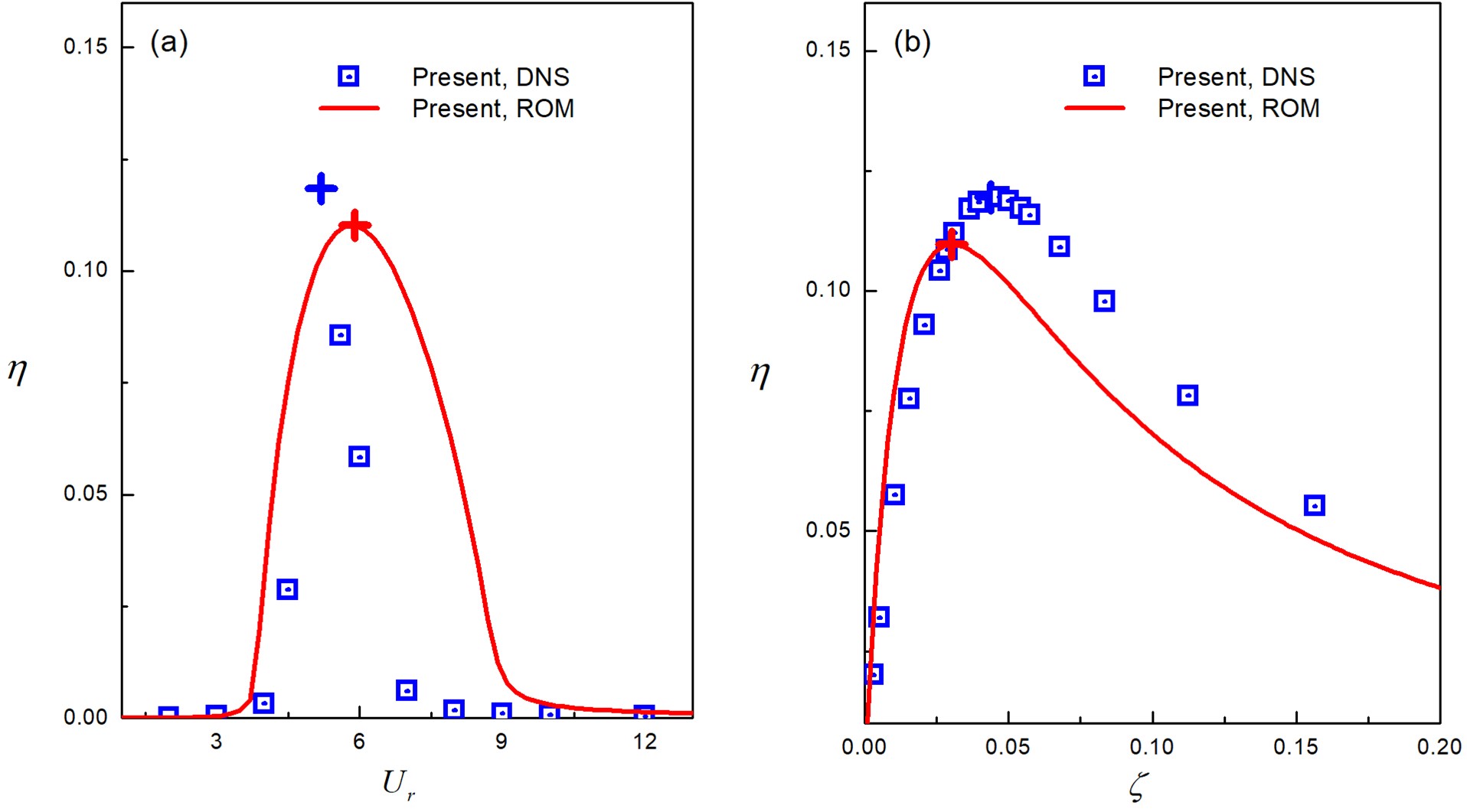}
	\caption{Comparisons of the variations of energy harvesting efficiency $\eta$ as a function of the (a) reduced velocity $U_{r}$ and (b) damping ratio $\zeta$, obtained by our ROM and quasi direct numerical simulations. "$+$" presents the maximum efficiency point.}
	\label{LowRe_Validation}
\end{figure}

In order to further support the results obtained by our ROM, we
validate them against the numerical data predicted by our DNS-FSI solver.
Figure~\ref{LowRe_Validation} shows the predicted $\eta$ as a function
of $U_{r}$ in $(a)$ and $\zeta$ in $(b)$. Note that in
figure~\ref{LowRe_Validation}$(a)$, the damping ratio is fixed at
$\zeta=0.044$, while in figure~\ref{LowRe_Validation}$(b)$ the reduced
velocity is fixed at $U_{r}=5.2$. Qualitative and to some extent
quantitative agreements can be found between our ROM and DNS results
in terms of the maximum efficiency locations, as well as the effects
of $U_r$ and $\zeta$. In addition, the maximum $\eta$ values predicted
by our ROM and DNS-FSI tests agree well with the numerical simulations
by \citet{Soti2017} who showed a maximum $\eta$ value of 0.13.

\section{Discussions} \label{Sec4}
\begin{figure}
	\centering
	\includegraphics[width=15cm]{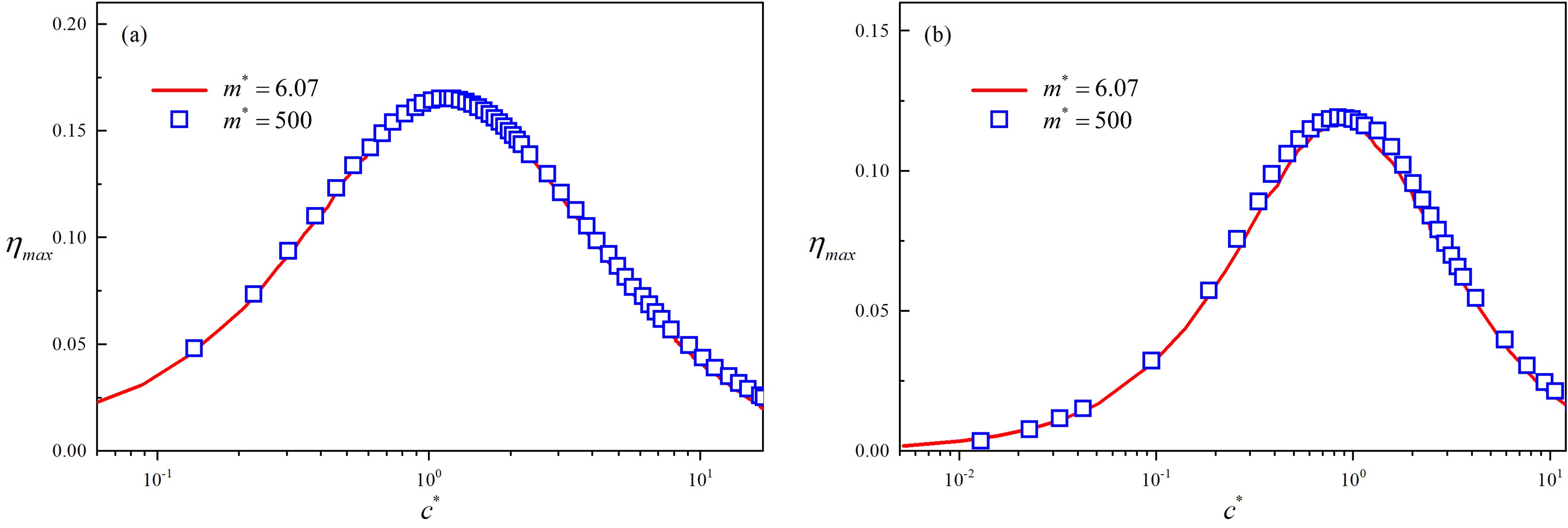}
	\caption{The variations of optimal energy harvesting
		efficiency as a function of the dimensionless parameter
		$c^{*}$ at (a) $Re = 6,000$ and (b) $Re = 150$.}
	\label{Eta_opt with Sc}
\end{figure}

From figures \ref{HighRe_Lowmass}, \ref{HighRe_Highmass} and
\ref{optimazition for low Re}, it is found that although the
energy harvesting characteristics for a heavy ($m^* = 500$) and
light ($m^* = 6$) VIV systems under flow are distinctly different,
the global maximum efficiency points are identical, regardless two
different Reynolds numbers in the present study. In addition, the
parameter $c^*=2C_s\omega_v/\rho U^2$ for the global optimal
efficiency point was found to be also nearly identical. The
above findings are further confirmed through the comparisons of the
optimal efficiency $\eta_{max}$ respect to the $c^*$ under
different mass ratios and Reynolds numbers, as shown in
figure~\ref{Eta_opt with Sc}. The comparisons support that not only the global optimal efficiency
but also the maximum efficiency under each damping ratio $\zeta$ are
governed by $c^*$ number. Interestingly, many previous
studies focusing on the effect of $c^*$ or combined
mass-damping parameter on the maximum amplitude of VIV have
indicated that the maximum amplitude can be expressed as a function of
the related controlled parameter by some empirical parameters
\citep[e.g.][]{Williamson2004,SotiJFS2018},
or more importantly by linearised reduced-order models
\citep[e.g.][]{Facchinetti2004}. This is interesting, because the
underlying relationship between the maximum amplitude and the
efficiency shows a certain way for future work to simplify the
relationship by linearising our ROM to describe the optimal energy
harvesting efficiency from VIV.

As can be seen again from figures \ref{HighRe_Lowmass},
\ref{HighRe_Highmass} and \ref{optimazition for low Re}, the optimal
energy harvesting efficiency is strongly related to Reynolds number,
indicating that the VIV energy harvester performs better in a
subcritical flow than that in a laminar flow. The results agrees well
with the findings from previous studies that the energy harvesting
efficiency tends to increase with Reynolds number. For example, the
numerical results of \citet{Soti2017} reported the energy harvesting
efficiency to be 0.10, 0.13 and 0.145 for $Re = 100$, $150$ and $200$,
respectively, while the experimental results of \citet{SotiJFS2018}
showed that the efficiency can be varied from 0.151 to 0.200 for
$10^3 < Re < 10^4$. From their experiments at
$Re \approx 7.5\times10^{4}$, \citet{LeeOE} reported a maximum
efficiency about 0.33. However, from the perspective of computations
of our proposed ROM, the differences between low- and high-$Re$ cases
are related to the lift coefficient ($C_{L0}$) and Strouhal number
($St$) of flow over a fixed circular cylinder. According to
\citet{Norberg2003}, $C_{L0}$ and $S_{t}$ do not monotonically vary
with $Re$; thus, the efficiency may not always increase with $Re$ as
an optimal Reynolds number would be expected to occur with a large
$C_{L0}$ and an appropriate $St$. Further investigations on this would be warranted.

\section{Conclusions} \label{Sec5}

We have applied a combined approach based on a reduced-order model, a DNS-FSI solver, and experiments to investigate the optimal efficiency of fluid-flow energy harvesting from VIV of a circular cylinder over a wide parametric space of reduced velocity and structural damping ratio. Based on the results, conclusions from the present work are drawn below.

The maximum efficiency and the optimal settings of
damping ratio and reduced velocity under different conditions were successfully predicted by the reduced-order model and validated against either our experiments and quasi-DNS. This indicates that the ROM is an low-cost tool for optimization of efficiency and it is valuable for related engineering designs. Moreover, via ROM, it shows that the maximum energy harvesting efficiency is strongly controlled by a dimensionless damping-mass parameter $c^*$ and the reduced velocity $U_r$. 

A VIV-energy converter works better in a subcritical flow than that in a laminar flow. Combing the present results and those in literature, it seems that a high Re would lead to a better efficiency in terms of energy harvesting. However, from the mechanics among the present reduced-order model, a large
$C_{L0}$ and an appropriate $St$ are the key parameters affecting the efficiency and therefore an optimal Re is expected. Further investigations on this would be warranted.

As it has been shown in the present work, a low-mass-ratio VIV energy
harvester performs better than a high-mass-ratio system, indicating
that an ocean energy harvester has more potential in fluid power
conversion than that of a wind energy harvester. This fact is because
the mass ratio effect plays an important role affecting the dynamics
in the lock-in region. As it has been reported in the literature
\citep[e.g.][]{Williamson2004}, the circular cylinder with an
extremely low mass ratio may undergo high-amplitude VIV over a larger range of reduced velocities, even to an infinite reduced velocity (achieved by setting $K=0$); thus, it would be of great interest to combine reduced-reduced modelling and experimental approaches to further investigate the energy harvesting performance from VIV of an extremely-low-mass-ratio body and, of course, other flow-induced vibration systems.


\section*{Acknowledgement}
\label{sec:acknowledgements}

This work was supported by the Australian Research Council through a
Discovery Early Career Researcher Award (DE200101650, J.Z.).


%
%

	%
	
	
	\bibliographystyle{jfm}
	\biboptions{authoryear}
	\bibliography{myref,references}

\begin{thebibliography}{43}
\expandafter\ifx\csname natexlab\endcsname\relax\def\natexlab#1{#1}\fi
\def\au#1{#1} \def\ed#1{#1} \def\yr#1{#1}\def\at#1{#1}\def\jt#1{\textit{#1}}
  \def\bt#1{#1}\def\bvol#1{\textbf{#1}} \def\vol#1{#1} \def\pg#1{#1}
  \def\publ#1{#1}\def\arxiv#1{#1}\def\org#1{#1}\def\st#1{\textit{#1}}

\bibitem[Abdelkefi {\em et~al.\/}(2012)Abdelkefi, Hajj \&
  Nayfeh]{Abdelkefi2012}
{\sc \au{Abdelkefi, A.}, \au{Hajj, M.~R.} \& \au{Nayfeh, A.~H.}} \yr{2012}
  \at{Phenomena and modeling of piezoelectric energy harvesting from freely
  oscillating cylinders}.  \jt{Nonlinear Dynamics}  \bvol{70}~(2),
  \pg{1377--1388}.

\bibitem[Bao {\em et~al.\/}(2012)Bao, Huang, Zhou, Tu \& Han]{Bao2012}
{\sc \au{Bao, Yan}, \au{Huang, Cheng}, \au{Zhou, Dai}, \au{Tu, Jiahuang} \&
  \au{Han, Zhaolong}} \yr{2012}  \at{Two-degree-of-freedom flow-induced
  vibrations on isolated and tandem cylinders with varying natural frequency
  ratios}.  \jt{Journal of Fluids and Structures}  \bvol{35},  \pg{50--75}.

\bibitem[Barrero-Gil {\em et~al.\/}(2012)Barrero-Gil, Pindado \&
  Avila]{Barrero-Gil2012}
{\sc \au{Barrero-Gil, Antonio}, \au{Pindado, Santiago} \& \au{Avila, Sergio}}
  \yr{2012}  \at{Extracting energy from vortex-induced vibrations: A parametric
  study}.  \jt{Applied Mathematical Modelling}  \bvol{36}~(7),
  \pg{3153--3160}.

\bibitem[Barrero-Gil {\em et~al.\/}(2009)Barrero-Gil, Sanz-Andrés \&
  Alonso]{Barrero-Gil2009a}
{\sc \au{Barrero-Gil, A.}, \au{Sanz-Andrés, A.} \& \au{Alonso, G.}} \yr{2009}
  \at{Hysteresis in transverse galloping: The role of the inflection points}.
  \jt{Journal of Fluids and Structures}  \bvol{25}~(6),  \pg{1007--1020}.

\bibitem[Bernitsas {\em et~al.\/}(2008)Bernitsas, Raghavan, Ben-Simon \&
  Garcia]{Bernitsas2008}
{\sc \au{Bernitsas, Michael~M.}, \au{Raghavan, Kamaldev}, \au{Ben-Simon, Y.} \&
  \au{Garcia, E.~M.}} \yr{2008}  \at{{VIVACE (Vortex Induced Vibration Aquatic
  Clean Energy)}: A new concept in generation of clean and renewable energy
  from fluid flow}.  \jt{Journal of Offshore Mechanics and Arctic Engineering}
  \bvol{130}~(4),  \pg{041101--041101--15}.

\bibitem[Blevins(2001)]{Blevins2001}
{\sc \au{Blevins, Robert~D.}} \yr{2001} {\em {Flow-Induced Vibration}\/}.
  \publ{Krieger Publishing Company}.

\bibitem[Ding {\em et~al.\/}(2015)Ding, Zhang, Wu, Mao \& Jiang]{Ding2015}
{\sc \au{Ding, Lin}, \au{Zhang, Li}, \au{Wu, Chunmei}, \au{Mao, Xinru} \&
  \au{Jiang, Deyi}} \yr{2015}  \at{Flow induced motion and energy harvesting of
  bluff bodies with different cross sections}.  \jt{Energy Conversion and
  Management}  \bvol{91},  \pg{416--426}.

\bibitem[Facchinetti {\em et~al.\/}(2004)Facchinetti, de~Langre \&
  Biolley]{Facchinetti2004}
{\sc \au{Facchinetti, M.~L.}, \au{de~Langre, E.} \& \au{Biolley, F.}} \yr{2004}
   \at{Coupling of structure and wake oscillators in vortex-induced
  vibrations}.  \jt{Journal of Fluids and Structures}  \bvol{19}~(2),
  \pg{123--140}.

\bibitem[Grouthier {\em et~al.\/}(2014)Grouthier, Michelin, Bourguet,
  Modarres-Sadeghi \& de~Langre]{Grouthier2014}
{\sc \au{Grouthier, Clément}, \au{Michelin, Sébastien}, \au{Bourguet, Rémi},
  \au{Modarres-Sadeghi, Yahya} \& \au{de~Langre, Emmanuel}} \yr{2014}  \at{On
  the efficiency of energy harvesting using vortex-induced vibrations of
  cables}.  \jt{Journal of Fluids and Structures}  \bvol{49},  \pg{427--440}.

\bibitem[Han {\em et~al.\/}(2021{\natexlab{{\em a\/}}})Han, Huang, Pan, Wang,
  Zhang \& Qin]{HanAIP}
{\sc \au{Han, Peng}, \au{Huang, Qiaogao}, \au{Pan, Guang}, \au{Wang, Wei},
  \au{Zhang, Tianqi} \& \au{Qin, Denghui}} \yr{2021{\natexlab{{\em a\/}}}}
  \at{Energy harvesting from flow-induced vibration of a low-mass square
  cylinder with different incidence angles}.  \jt{AIP Advances}  \bvol{11}~(2),
   \pg{025126}.

\bibitem[Han {\em et~al.\/}(2021{\natexlab{{\em b\/}}})Han, Hémon, Pan \&
  de~Langre]{HanND}
{\sc \au{Han, Peng}, \au{Hémon, Pascal}, \au{Pan, Guang} \& \au{de~Langre,
  Emmanuel}} \yr{2021{\natexlab{{\em b\/}}}}  \at{Nonlinear modeling of
  combined galloping and vortex-induced vibration of square sections under
  flow}.  \jt{Nonlinear Dynamics}  \bvol{103}~(4),  \pg{3113--3125}.

\bibitem[Han \& de~Langre(2022)]{Han2021JFM}
{\sc \au{Han, Peng} \& \au{de~Langre, Emmanuel}} \yr{2022}  \at{There is no
  critical mass ratio for galloping of a square cylinder under flow}.
  \jt{Journal of Fluid Mechanics}  \bvol{931},  \pg{A27}.

\bibitem[Han {\em et~al.\/}(2018)Han, Pan \& Tian]{HanPOF}
{\sc \au{Han, Peng}, \au{Pan, Guang} \& \au{Tian, Wenlong}} \yr{2018}
  \at{Numerical simulation of flow-induced motion of three rigidly coupled
  cylinders in equilateral-triangle arrangement}.  \jt{Physics of Fluids}
  \bvol{30}~(12),  \pg{125107}.

\bibitem[Han {\em et~al.\/}(2020)Han, Pan, Zhang, Wang \& Tian]{HanOE}
{\sc \au{Han, Peng}, \au{Pan, Guang}, \au{Zhang, Baoshou}, \au{Wang, Wei} \&
  \au{Tian, Wenlong}} \yr{2020}  \at{Three-cylinder oscillator under flow: Flow
  induced vibration and energy harvesting}.  \jt{Ocean Engineering}
  \bvol{211},  \pg{107619}.

\bibitem[Kandasamy {\em et~al.\/}(2016)Kandasamy, Cui, Townsend, Foo, Guo,
  Shenoi \& Xiong]{Kandasamy2016}
{\sc \au{Kandasamy, Ramkumar}, \au{Cui, Fangsen}, \au{Townsend, Nicholas},
  \au{Foo, Choon~Chiang}, \au{Guo, Junyan}, \au{Shenoi, Ajit} \& \au{Xiong,
  Yeping}} \yr{2016}  \at{A review of vibration control methods for marine
  offshore structures}.  \jt{Ocean Engineering}  \bvol{127},  \pg{279--297}.

\bibitem[Lai {\em et~al.\/}(2021)Lai, Wang, Zhu, Zhang, Wang, Yang \&
  Yurchenko]{Lai2021MSSP}
{\sc \au{Lai, Zhihui}, \au{Wang, Shuaibo}, \au{Zhu, Likuan}, \au{Zhang,
  Guoqing}, \au{Wang, Junlei}, \au{Yang, Kai} \& \au{Yurchenko, Daniil}}
  \yr{2021}  \at{A hybrid piezo-dielectric wind energy harvester for
  high-performance vortex-induced vibration energy harvesting}.  \jt{Mechanical
  Systems and Signal Processing}  \bvol{150},  \pg{107212}.

\bibitem[de~Langre(2006)]{deLangre2006}
{\sc \au{de~Langre, E.}} \yr{2006}  \at{Frequency lock-in is caused by
  coupled-mode flutter}.  \jt{Journal of Fluids and Structures}
  \bvol{22}~(6-7),  \pg{783--791}.

\bibitem[Lee \& Bernitsas(2011)]{LeeOE}
{\sc \au{Lee, J.~H.} \& \au{Bernitsas, M.~M.}} \yr{2011}  \at{{High-damping,
  high-Reynolds VIV tests for energy harnessing using the VIVACE converter}}.
  \jt{Ocean Engineering}  \bvol{38}~(16),  \pg{1697--1712}.

\bibitem[Ma {\em et~al.\/}(2016)Ma, Sun, Nowakowski, Mauer \&
  Bernitsas]{Ma2016}
{\sc \au{Ma, Chunhui}, \au{Sun, Hai}, \au{Nowakowski, Gary}, \au{Mauer, Erik}
  \& \au{Bernitsas, Michael~M.}} \yr{2016}  \at{Nonlinear piecewise restoring
  force in hydrokinetic power conversion using flow induced motions of single
  cylinder}.  \jt{Ocean Engineering}  \bvol{128},  \pg{1--12}.

\bibitem[Mishra {\em et~al.\/}(2020)Mishra, Soti, Bhardwaj, Kulkarni \&
  Thompson]{Mishra2020}
{\sc \au{Mishra, Rahul}, \au{Soti, Atul}, \au{Bhardwaj, Rajneesh},
  \au{Kulkarni, Salil~S.} \& \au{Thompson, Mark~C.}} \yr{2020}  \at{Transverse
  vortex-induced vibration of a circular cylinder on a viscoelastic support at
  low reynolds number}.  \jt{Journal of Fluids and Structures}  \bvol{95},
  \pg{102997}.

\bibitem[Norberg(2003)]{Norberg2003}
{\sc \au{Norberg, C.}} \yr{2003}  \at{Fluctuating lift on a circular cylinder:
  review and new measurements}.  \jt{Journal of Fluids and Structures}
  \bvol{17}~(1),  \pg{57--96}.

\bibitem[Païdoussis {\em et~al.\/}(2010)Païdoussis, Price \&
  De~Langre]{Paidoussis2010}
{\sc \au{Païdoussis, Michael~P.}, \au{Price, Stuart~J.} \& \au{De~Langre,
  Emmanuel}} \yr{2010} {\em {Fluid-Structure Interactions: Cross-Flow-Induced
  Instabilities}\/}.  \publ{Cambridge University Press}.

\bibitem[Qu {\em et~al.\/}(2013)Qu, Norberg, Davidson, Peng \& Wang]{Qu2013}
{\sc \au{Qu, Lixia}, \au{Norberg, Christoffer}, \au{Davidson, Lars}, \au{Peng,
  Shia-Hui} \& \au{Wang, Fujun}} \yr{2013}  \at{Quantitative numerical analysis
  of flow past a circular cylinder at reynolds number between 50 and 200}.
  \jt{Journal of Fluids and Structures}  \bvol{39},  \pg{347--370}.

\bibitem[Rostami \& Armandei(2017)]{RostamiRSER2017}
{\sc \au{Rostami, Ali~Bakhshandeh} \& \au{Armandei, Mohammadmehdi}} \yr{2017}
  \at{Renewable energy harvesting by vortex-induced motions: Review and
  benchmarking of technologies}.  \jt{Renewable and Sustainable Energy Reviews}
   \bvol{70},  \pg{193--214}.

\bibitem[Sarpkaya(2004)]{Sarpkaya2004}
{\sc \au{Sarpkaya, T.}} \yr{2004}  \at{A critical review of the intrinsic
  nature of vortex-induced vibrations}.  \jt{Journal of Fluids and Structures}
  \bvol{19}~(4),  \pg{389--447}.

\bibitem[Soti {\em et~al.\/}(2017)Soti, Thompson, Sheridan \&
  Bhardwaj]{Soti2017}
{\sc \au{Soti, Atul~Kumar}, \au{Thompson, Mark~C.}, \au{Sheridan, John} \&
  \au{Bhardwaj, Rajneesh}} \yr{2017}  \at{Harnessing electrical power from
  vortex-induced vibration of a circular cylinder}.  \jt{Journal of Fluids and
  Structures}  \bvol{70},  \pg{360--373}.

\bibitem[Soti {\em et~al.\/}(2018)Soti, Zhao, Thompson, Sheridan \&
  Bhardwaj]{SotiJFS2018}
{\sc \au{Soti, A.~K.}, \au{Zhao, J.~S.}, \au{Thompson, M.~C.}, \au{Sheridan,
  J.} \& \au{Bhardwaj, R.}} \yr{2018}  \at{Damping effects on vortex-induced
  vibration of a circular cylinder and implications for power extraction}.
  \jt{Journal of Fluids and Structures}  \bvol{81},  \pg{289--308}.

\bibitem[Srinil \& Zanganeh(2012)]{Srinil2012}
{\sc \au{Srinil, Narakorn} \& \au{Zanganeh, Hossein}} \yr{2012}  \at{{Modelling
  of coupled cross-flow/in-line vortex-induced vibrations using double Duffing
  and van der Pol oscillators}}.  \jt{Ocean Engineering}  \bvol{53},
  \pg{83--97}.

\bibitem[Vandiver(2012)]{Vandiver2012}
{\sc \au{Vandiver, J.~Kim}} \yr{2012}  \at{Damping parameters for flow-induced
  vibration}.  \jt{Journal of Fluids and Structures}  \bvol{35},
  \pg{105--119}.

\bibitem[Violette {\em et~al.\/}(2007)Violette, de~Langre \&
  Szydlowski]{Violette2007}
{\sc \au{Violette, R.}, \au{de~Langre, E.} \& \au{Szydlowski, J.}} \yr{2007}
  \at{Computation of vortex-induced vibrations of long structures using a wake
  oscillator model: Comparison with {DNS} and experiments}.  \jt{Computers and
  Structures}  \bvol{85}~(11-14),  \pg{1134--1141}.

\bibitem[Violette {\em et~al.\/}(2010)Violette, de~Langre \&
  Szydlowski]{Violette2010}
{\sc \au{Violette, R.}, \au{de~Langre, E.} \& \au{Szydlowski, J.}} \yr{2010}
  \at{A linear stability approach to vortex-induced vibrations and waves}.
  \jt{Journal of Fluids and Structures}  \bvol{26}~(3),  \pg{442--466}.

\bibitem[Wang {\em et~al.\/}(2020{\natexlab{{\em a\/}}})Wang, Geng, Ding, Zhu
  \& Yurchenko]{Wang2020AE}
{\sc \au{Wang, Junlei}, \au{Geng, Linfeng}, \au{Ding, Lin}, \au{Zhu, Hongjun}
  \& \au{Yurchenko, Daniil}} \yr{2020{\natexlab{{\em a\/}}}}  \at{The
  state-of-the-art review on energy harvesting from flow-induced vibrations}.
  \jt{Applied Energy}  \bvol{267},  \pg{114902}.

\bibitem[Wang {\em et~al.\/}(2021)Wang, Yurchenko, Hu, Zhao, Tang \&
  Yang]{Wang2021APL}
{\sc \au{Wang, Junlei}, \au{Yurchenko, Daniil}, \au{Hu, Guobiao}, \au{Zhao,
  Liya}, \au{Tang, Lihua} \& \au{Yang, Yaowen}} \yr{2021}  \at{Perspectives in
  flow-induced vibration energy harvesting}.  \jt{Applied Physics Letters}
  \bvol{119}~(10).

\bibitem[Wang {\em et~al.\/}(2019)Wang, Zhou, Zhang \& Yurchenko]{WangECM2019}
{\sc \au{Wang, Junlei}, \au{Zhou, Shengxi}, \au{Zhang, Zhien} \& \au{Yurchenko,
  Daniil}} \yr{2019}  \at{High-performance piezoelectric wind energy harvester
  with {Y}-shaped attachments}.  \jt{Energy Conversion and Management}
  \bvol{181},  \pg{645--652}.

\bibitem[Wang {\em et~al.\/}(2020{\natexlab{{\em b\/}}})Wang, Song, Mao, Tian,
  Zhang \& Han]{Wang2020}
{\sc \au{Wang, Wei}, \au{Song, Baowei}, \au{Mao, Zhaoyong}, \au{Tian, Wenlong},
  \au{Zhang, Tingying} \& \au{Han, Peng}} \yr{2020{\natexlab{{\em b\/}}}}
  \at{Numerical investigation on vortex-induced vibration of bluff bodies with
  different rear edges}.  \jt{Ocean Engineering}  \bvol{197},  \pg{106871}.

\bibitem[Williamson \& Govardhan(2004)]{Williamson2004}
{\sc \au{Williamson, C. H.~K.} \& \au{Govardhan, R.}} \yr{2004}
  \at{Vortex-induced vibrations}.  \jt{Annual Review of Fluid Mechanics}
  \bvol{36}~(1),  \pg{413--455}.

\bibitem[Wong {\em et~al.\/}(2018)Wong, Zhao, {Lo Jacono}, Thompson \&
  Sheridan]{wong2018}
{\sc \au{Wong, K.~W.~L.}, \au{Zhao, J.}, \au{{Lo Jacono}, D.}, \au{Thompson,
  M.~C.} \& \au{Sheridan, J.}} \yr{2018}  \at{{Experimental investigation of
  flow-induced vibration of a sinusoidally rotating circular cylinder}}.
  \jt{Journal of Fluid Mechanics}  \bvol{848},  \pg{430--466}.

\bibitem[Zanganeh \& Srinil(2016)]{Zanganeh2016}
{\sc \au{Zanganeh, Hossein} \& \au{Srinil, Narakorn}} \yr{2016}
  \at{Three-dimensional {VIV} prediction model for a long flexible cylinder
  with axial dynamics and mean drag magnifications}.  \jt{Journal of Fluids and
  Structures}  \bvol{66},  \pg{127--146}.

\bibitem[Zhao {\em et~al.\/}(2018{\natexlab{{\em a\/}}})Zhao, Hourigan \&
  Thompson]{zhao2018Dsection}
{\sc \au{Zhao, J.}, \au{Hourigan, K.} \& \au{Thompson, M.~C.}}
  \yr{2018{\natexlab{{\em a\/}}}}  \at{{Flow-induced vibration of D-section
  cylinders: an afterbody is not essential for vortex-induced vibration}}.
  \jt{Journal of Fluid Mechanics}  \bvol{851},  \pg{317--343}.

\bibitem[Zhao {\em et~al.\/}(2022)Zhao, Hourigan \& Thompson]{zhao2022prf}
{\sc \au{Zhao, J.}, \au{Hourigan, K.} \& \au{Thompson, M.~C.}} \yr{2022}
  \at{{Damping effect on transverse flow-induced vibration of a rotating
  circular cylinder and its implied energy harvesting performance}}.
  \jt{Physical Review Fluids}  \bvol{7}.

\bibitem[Zhao {\em et~al.\/}(2014)Zhao, Leontini, {Lo Jacono} \&
  Sheridan]{zhao2014fsi}
{\sc \au{Zhao, J.}, \au{Leontini, J.~S.}, \au{{Lo Jacono}, D.} \& \au{Sheridan,
  J.}} \yr{2014}  \at{Fluid--structure interaction of a square cylinder at
  different angles of attack}.  \jt{Journal of Fluid Mechanics}  \bvol{747},
  \pg{688--721}.

\bibitem[Zhao {\em et~al.\/}(2018{\natexlab{{\em b\/}}})Zhao, {Lo Jacono},
  Sheridan, Hourigan \& Thompson]{zhao2018inline}
{\sc \au{Zhao, J.}, \au{{Lo Jacono}, D.}, \au{Sheridan, J.}, \au{Hourigan, K.}
  \& \au{Thompson, M.~C.}} \yr{2018{\natexlab{{\em b\/}}}}  \at{Experimental
  investigation of in-line flow-induced vibration of a rotating cylinder}.
  \jt{Journal of Fluid Mechanics}  \bvol{847},  \pg{664--699}.

\bibitem[Zhao(2013)]{Zhao2013}
{\sc \au{Zhao, Ming}} \yr{2013}  \at{Flow induced vibration of two rigidly
  coupled circular cylinders in tandem and side-by-side arrangements at a low
  reynolds number of 150}.  \jt{Physics of Fluids}  \bvol{25}~(12),
  \pg{123601}.

\end{thebibliography}
	
\end{document}